\documentclass[sigconf]{acmart}
\AtBeginDocument{%
  }

\usepackage{algorithm}
\usepackage{algpseudocode}
\usepackage{subcaption}
\usepackage{tabularx}
\usepackage{array}
\usepackage{siunitx}
\usepackage{booktabs}
\usepackage{multirow}
\newcolumntype{Y}{>{\raggedright\arraybackslash}X}
\renewcommand\footnotetextcopyrightpermission[1]{}

\setcopyright{none}
\acmDOI{}
\acmISBN{}
\acmConference{}{}{}

\usepackage{tikz}
\usepackage{adjustbox}
\usetikzlibrary{positioning, shapes.geometric, arrows.meta, calc, fit, backgrounds}

\definecolor{csblue}{RGB}{235, 245, 255}
\definecolor{csbluedark}{RGB}{50, 120, 200}
\definecolor{csorange}{RGB}{255, 240, 230}
\definecolor{csorangedark}{RGB}{200, 100, 30}
\definecolor{cspurple}{RGB}{245, 235, 255}
\definecolor{cspurpledark}{RGB}{120, 50, 200}
\definecolor{csgreen}{RGB}{235, 250, 235}
\definecolor{csgreendark}{RGB}{40, 160, 40}
\definecolor{csred}{RGB}{255, 235, 235}
\definecolor{csreddark}{RGB}{200, 50, 50}
\definecolor{csyellow}{RGB}{255, 250, 220}
\definecolor{csyellowdark}{RGB}{180, 150, 0}

\begin{document}

\title[Agentic-Kube: A Graph-Enhanced Multi-Agent Reinforcement Learning Framework for Multi-Objective Kubernetes Scheduling]{Agentic-Kube: A Graph-Enhanced Multi-Agent Reinforcement Learning Framework for Multi-Objective Kubernetes Scheduling}

\author{Hamed Hamzeh}
\email{H.Hamzeh@westminster.ac.uk}
\affiliation{%
  \institution{University of Westminster}
  \department{Computer Science and Engineering}
  \city{London}
  \country{United Kingdom}}

\renewcommand{\shortauthors}{Hamzeh}

\begin{abstract}
Cloud-native container orchestration requires resource schedulers capable of balancing infrastructure expenditure, fault resilience, and node utilisation. Conventional reinforcement learning approaches typically rely on monolithic single-agent models that suffer from gradient interference and reward dilution when mapping conflicting operational goals into a single scalar reward. We present Agentic-Kube, a cooperative multi-agent reinforcement learning framework designed for real-time Kubernetes pod placement. The architecture decomposes multi-objective scheduling into a tripartite optimisation space managed by dedicated sub-agents for cost minimisation, anti-affinity fault tolerance, and vector resource balancing. Agentic-Kube integrates a bipartite Graph Convolutional Network to capture dynamic host-pod dependencies, a two-stage monotonic QMIX value factorisation network to maintain joint action value coherence, and a plurality voting consensus mechanism with action feasibility masking against allocatable node predicates. We evaluate the framework across live heterogeneous Google Kubernetes Engine deployments and macro-scale cluster environments spanning 50 to 1,000 nodes under empirical Alibaba trace data, diurnal microservice variations, and flash-crowd bursts. Across physical and simulated evaluations against native Kubernetes heuristics, Single-Agent PPO, Multi-Agent PPO, and evolutionary NSGA-II bounds, Agentic-Kube consistently achieves Pareto-efficient placements. In diurnal microservice workloads, it reduces anti-affinity service collisions to 7.11\%, representing a 53.0\% relative reduction compared to the default Kubernetes scheduler. Under Alibaba traces, the policy achieves a 65.15\% spot instance allocation ratio, while macro-scale benchmarks demonstrate scaling up to 1,000 nodes with mean decision latencies under 17\,ms and 99th-percentile latencies under 31\,ms, executing without container restart failures and operating well within standard scheduling admission timeouts.
\end{abstract}

\keywords{Cloud Computing, Centralised training, Decentralised execution, Graph Neural Networks (GNNs), Kubernetes, Multi-Agent Reinforcement Learning (MARL), Orchestration, Scheduling}

\maketitle

\section{Introduction}
Kubernetes serves as the de facto open-source container orchestration platform for dynamic, distributed cloud applications~\cite{b1,b2}. The fundamental deployment unit in Kubernetes is the \textit{pod}, representing an instance of a running process that encapsulates one or more tightly coupled containers~\cite{b3}. Assigning incoming pods to cluster nodes constitutes a multi-dimensional optimisation problem~\cite{b5}. Production placement routines must dynamically account for continuously fluctuating parameters, including heterogeneous workload profiles, node resource capacities, and competing operational objectives~\cite{b4}. Consequently, pod scheduling directly influences system stability, resource utilisation, infrastructure expenditure, and service responsiveness~\cite{b555}. The default Kubernetes scheduler employs filtering and scoring heuristics that evaluate basic allocatable capacities, but this mechanism struggles to manage complex multi-tenant environments with non-linear constraints~\cite{b777}. While specialised rule-based extensions have been proposed, they increase system maintenance complexity~\cite{b4,carmen} and exhibit rigid scalability boundaries under rapid workload distribution shifts~\cite{nicola,b4,carmen}.

To overcome the limitations of static algorithms, recent research has explored adaptive scheduling based on Reinforcement Learning (RL)~\cite{wang,b10c,b7,e4,e5}. Given the high-dimensional state space of modern cloud clusters, Deep Reinforcement Learning (DRL) is frequently applied to parameterise value and policy functions~\cite{b11}. DRL models the scheduler as an agent that acquires placement policies through continuous interaction with the cluster environment~\cite{b10c,wang,b7b,b10d,b10e,b10f,b10g,b10h,b10i,b47,e5,b17}. The agent observes cluster state representations, such as node resource utilisation and incoming pod specifications, and outputs placement actions guided by reward feedback. This data-driven approach enables the discovery of non-linear placement policies tailored to target operational environments~\cite{b7}.

However, conventional single-agent DRL formulations present structural limitations in large-scale cluster environments. Monolithic centralised agents suffer from poor dimensional scalability as the state-action space expands with cluster size~\cite{amanda,b777}. Centralised inference can introduce latency bottlenecks~\cite{wang}, create single points of failure, and struggle to converge to globally balanced multi-objective policies~\cite{zhou,jalali}. Multi-Agent Reinforcement Learning (MARL) provides a modular alternative by distributing decision responsibilities across specialised agents~\cite{park,b10b}. Nevertheless, multi-agent frameworks introduce coordination challenges~\cite{aina}, particularly when balancing competing operational metrics such as cost minimisation, fault tolerance, and multi-dimensional resource packing~\cite{b7a,farid,zhong}. Prior studies predominantly rely on static, linear scalar reward formulations to aggregate these competing goals~\cite{wang,e5,gaon}. This approach induces reward dilution and gradient interference, causing the network to over-index on dominant sub-rewards and collapse into sub-optimal greedy heuristics under severe resource contention.

To resolve these challenges, we introduce Agentic-Kube, a graph-enhanced multi-agent reinforcement learning framework for Kubernetes scheduling. Agentic-Kube decouples the multi-objective placement problem into a cooperative multi-agent environment governed by three specialised functional actors: a cost reduction agent, an anti-affinity fault tolerance agent, and a vector resource utilisation agent. The framework adopts a Centralised Training with Decentralised Execution (CTDE) paradigm~\cite{CTDE}. To prevent reward dilution and ensure coordinated policy convergence, Agentic-Kube implements a two-stage monotonic QMIX value factorisation network that guarantees joint action-value coherence. During inference, agents process topological embeddings generated by a bipartite Graph Convolutional Network (GCN) that explicitly models container-to-host graph dependencies, enabling spatial awareness without runtime inter-agent communication overhead. Individual agent proposals are reconciled through a plurality voting consensus mechanism constrained by strict action feasibility masking against Kubernetes allocatable predicates.

We evaluate Agentic-Kube across physical deployments on an asymmetric Google Kubernetes Engine (GKE) cluster and a macro-scale cluster simulation testbed scaling from 50 to 1,000 physical nodes. The evaluation benchmarks Agentic-Kube against the native Kubernetes scheduler, Single-Agent PPO, Multi-Agent PPO (MAPPO), and an evolutionary Pareto upper bound (NSGA-II) across three operational regimes: empirical Alibaba cluster traces, diurnal microservice variations, and flash-crowd bursts.

The primary contributions of this paper are summarised as follows:
\begin{enumerate}
    \item We formulate a tripartite multi-agent reinforcement learning architecture that mathematically decomposes Kubernetes scheduling into dedicated sub-agents for cost, fault tolerance, and vector utilisation, eliminating the reward dilution and gradient interference inherent in scalar single-agent DRL.
    \item We implement a two-stage monotonic QMIX value factorisation network within a Centralised Training with Decentralised Execution framework, guaranteeing coherent multi-agent joint action evaluation while supporting scalable inference.
    \item We design a bipartite Graph Convolutional Network (GCN) encoder to construct permutation-invariant topological state embeddings of active pods and physical nodes, providing graph-level structural awareness without requiring run-time inter-agent communication.
    \item We incorporate an action-masked plurality voting consensus mechanism that reconciles agent decisions in real time while strictly filtering out nodes that violate physical allocatable resource predicates.
    \item We conduct extensive empirical evaluations across a live 8-node GKE testbed and macro-scale cluster environments scaling to 1,000 nodes, demonstrating that Agentic-Kube achieves Pareto-dominant trade-offs, reduces anti-affinity collisions by up to 53.0\%, increases spot instance harvesting to 73.65\% (offline) and 91.53\% (online adaptive) at scale, and maintains sub-25\,ms decision latencies with zero container restart failures.
\end{enumerate}

The remainder of this paper is organised as follows. Section~2 reviews related work. Section~3 presents the problem formulation and AGMARL-DKS approach, while Section~4 describes the system design. Section~5 explains the experimental setup and workload generation, Section~6 presents the results, and Section~7 concludes the paper.

\section{Related Work}
\label{sec:related_work}

The application of machine learning to cloud resource orchestration has evolved from static heuristic tuning to adaptive, data-driven decision frameworks. Existing literature in reinforcement learning for workload placement can be broadly classified into three categories: monolithic single-agent deep reinforcement learning (DRL), distributed multi-agent reinforcement learning (MARL), and graph-structured representation models.

\subsection{Single-Agent Deep Reinforcement Learning Schedulers}
Early adaptations of reinforcement learning to cloud scheduling formulate pod or task placement as a Markov Decision Process (MDP) governed by a single centralised agent. The framework proposed in \cite{b10c} integrates deep neural networks with reinforcement learning to automate task scheduling and improve aggregate cluster utilisation. Similarly, Deep Q-Network (DQN) formulations have been developed for multi-stage workflow placement \cite{b10i}, while plugins such as RLKube \cite{e5} embed tabular and deep policy models directly within Kubernetes admission controllers. 

While these single-agent systems adapt to non-linear workload patterns better than static rule-based schedulers, they face severe structural constraints when deployed in enterprise clusters. Monolithic architectures depend on flattened state vectors that scale linearly with the number of physical nodes $|V|$. As cluster scale expands from dozens to hundreds of nodes, the resulting state-action space explosion induces dimensional collapse and high inference latency. Furthermore, these architectures compress distinct operational priorities into a single scalar reward function. Under severe resource contention, this linear scalarisation triggers reward dilution and gradient interference, causing the policy to overfit to dominant resource metrics while neglecting secondary constraints such as anti-affinity rules or spot pricing discounts.

\subsection{Multi-Agent Systems and Decentralised Scheduling}
To address the computational bottlenecks of centralised agents, distributed and multi-agent reinforcement learning architectures have been explored across cloud and edge computing. The DRL4HFC framework \cite{b10d} introduces a multi-agent model for microservice placement in hierarchical fog environments, minimising end-to-end latency and compute consumption. In IoT and edge paradigms, heuristic-guided multi-agent schemes such as the Dynamic Mayfly Optimisation Scheduling algorithm \cite{b10e} and spatial game-theoretic models based on cellular automata \cite{b10g} have been developed to balance regional workloads. 

Decentralised scheduling architectures \cite{b10f} improve system resilience by eliminating single points of failure, but they constrain agent decisions to localised state observations. This local perceptual boundary prevents individual nodes from assessing cluster-wide placement trade-offs. The Hierarchical Multi-Agent Optimisation (HMAO) algorithm \cite{b10h} attempts to coordinate sub-agents hierarchically, yet it lacks a mechanism for real-time global topology tracking. While multi-agent container allocation frameworks \cite{b47} demonstrate the feasibility of distributed decision-making in containerised environments, they lack formal value-mixing guarantees to coordinate agents with conflicting goals during resource starvation. 

Agentic-Kube departs from these methods by implementing a Centralised Training with Decentralised Execution (CTDE) paradigm. It decouples the scheduling problem into dedicated sub-agents representing cost minimisation, fault tolerance, and vector resource balancing. Rather than relying on uncoordinated actor policies or linear reward combinations, Agentic-Kube coordinates agent learning through a two-stage monotonic QMIX value factorisation network, guaranteeing mathematically stable joint action evaluation.

\subsection{Graph Representation and Multi-Objective Trade-Offs}
Capturing spatial dependencies and network topology is critical for placement decisions in distributed environments. Graph Neural Networks (GNNs) have recently been applied to cloud scheduling to generate permutation-invariant representations of computing clusters. The DRL-GIN framework \cite{b17} leverages Graph Isomorphism Networks to capture application topology for multi-objective microservice placement within a standard actor-critic architecture. However, existing GNN-based schedulers typically learn multi-objective trade-offs implicitly through shared policy layers, which remains susceptible to gradient conflict when objectives diverge sharply.

Agentic-Kube addresses these limitations by integrating a bipartite Graph Convolutional Network (GCN) encoder that explicitly constructs graph embeddings over physical nodes and active container instances. These topological embeddings are distributed across the specialised sub-agents without requiring inter-agent communication during runtime inference. Final scheduling actions are reconciled through an action-masked plurality voting consensus mechanism that strictly enforces physical Kubernetes predicate filters. Table~\ref{tab:comprehensive_comparison} provides a structural feature comparison between Agentic-Kube and prominent baseline architectures.

\begin{table*}[htbp]
\centering
\small
\caption{Architectural and functional comparison of cloud and edge scheduling frameworks.}
\label{tab:comprehensive_comparison}
\begin{tabular*}{\textwidth}{@{\extracolsep{\fill}}lcccccccc@{}}
\hline
\textbf{Framework} & \textbf{Multi-Agent} & \textbf{Graph} & \textbf{Monotonic Value} & \textbf{Kubernetes} & \textbf{Spot Cost} & \textbf{Topological} & \textbf{Action} & \textbf{Sub-25\,ms} \\
& \textbf{(CTDE)} & \textbf{Neural Nets} & \textbf{Factorisation} & \textbf{Native} & \textbf{Optimisation} & \textbf{Resilience} & \textbf{Masking} & \textbf{Latency} \\
\hline
DRL4HFC \cite{b10d} & $\checkmark$ & $\times$ & $\times$ & $\times$ & $\times$ & $\times$ & $\times$ & $\checkmark$ \\
DMOS \cite{b10e} & $\times$ & $\times$ & $\times$ & $\times$ & $\times$ & $\checkmark$ & $\times$ & $\times$ \\
Game-Theoretic \cite{b10g} & $\checkmark$ & $\times$ & $\times$ & $\times$ & $\times$ & $\checkmark$ & $\times$ & $\times$ \\
HMAO \cite{b10h} & $\checkmark$ & $\times$ & $\times$ & $\times$ & $\times$ & $\times$ & $\times$ & $\checkmark$ \\
RLKube \cite{e5} & $\times$ & $\times$ & $\times$ & $\checkmark$ & $\times$ & $\times$ & $\checkmark$ & $\checkmark$ \\
DRL Workflow \cite{b10i} & $\times$ & $\times$ & $\times$ & $\times$ & $\times$ & $\times$ & $\times$ & $\times$ \\
DRL-GIN \cite{b17} & $\times$ & $\checkmark$ & $\times$ & $\times$ & $\times$ & $\checkmark$ & $\times$ & $\times$ \\
Multi-Agent DRL \cite{b47} & $\checkmark$ & $\times$ & $\times$ & $\checkmark$ & $\times$ & $\times$ & $\times$ & $\checkmark$ \\
\textbf{Agentic-Kube (Ours)} & $\mathbf{\checkmark}$ & $\mathbf{\checkmark}$ & $\mathbf{\checkmark}$ & $\mathbf{\checkmark}$ & $\mathbf{\checkmark}$ & $\mathbf{\checkmark}$ & $\mathbf{\checkmark}$ & $\mathbf{\checkmark}$ \\
\hline
\end{tabular*}
\end{table*}

\section{Problem Formulation}
\label{sec:problem_formulation}

The scheduling of pods within a Kubernetes cluster is modelled as a sequential decision-making process under resource and topological constraints. We define the physical cluster as a graph $\mathcal{G} = (\mathcal{V}, \mathcal{E})$, where the vertices $\mathcal{V} = \{v_1, v_2, \dots, v_N\}$ represent the available worker nodes, and the edges $\mathcal{E}$ represent the network topology interconnecting them. 

Each node $v_n \in \mathcal{V}$ possesses a dynamic state vector $s_n(t) = [C_n(t), M_n(t), T_n, A_n(t)]$, where $C_n(t)$ and $M_n(t)$ denote the available CPU and memory capacity at time $t$. The variable $T_n \in \{0, 1\}$ indicates the hardware tier, where $1$ represents a transient spot instance and $0$ represents a predictable on-demand instance. The component $A_n(t)$ represents the active allocation set, defined as the collection of service identifiers $\{l_1, l_2, \dots, l_m\}$ for all pods currently executing on node $v_n$. This precise tracking is mathematically required to compute topological collisions and enforce microservice anti-affinity constraints.

An incoming scheduling request is defined as a pod $p_k = [c_k, m_k, l_k]$, requiring specific CPU limits $c_k$, memory limits $m_k$, and possessing a service identifier $l_k$. The objective of the scheduler is to map $p_k$ to a node $v_n$ such that operational requirements are satisfied across three conflicting dimensions: financial cost, system fault tolerance, and spatial resource utilisation~\cite{farid}.

\begin{figure*}[htbp]
    \centering
    \includegraphics[width=\textwidth]{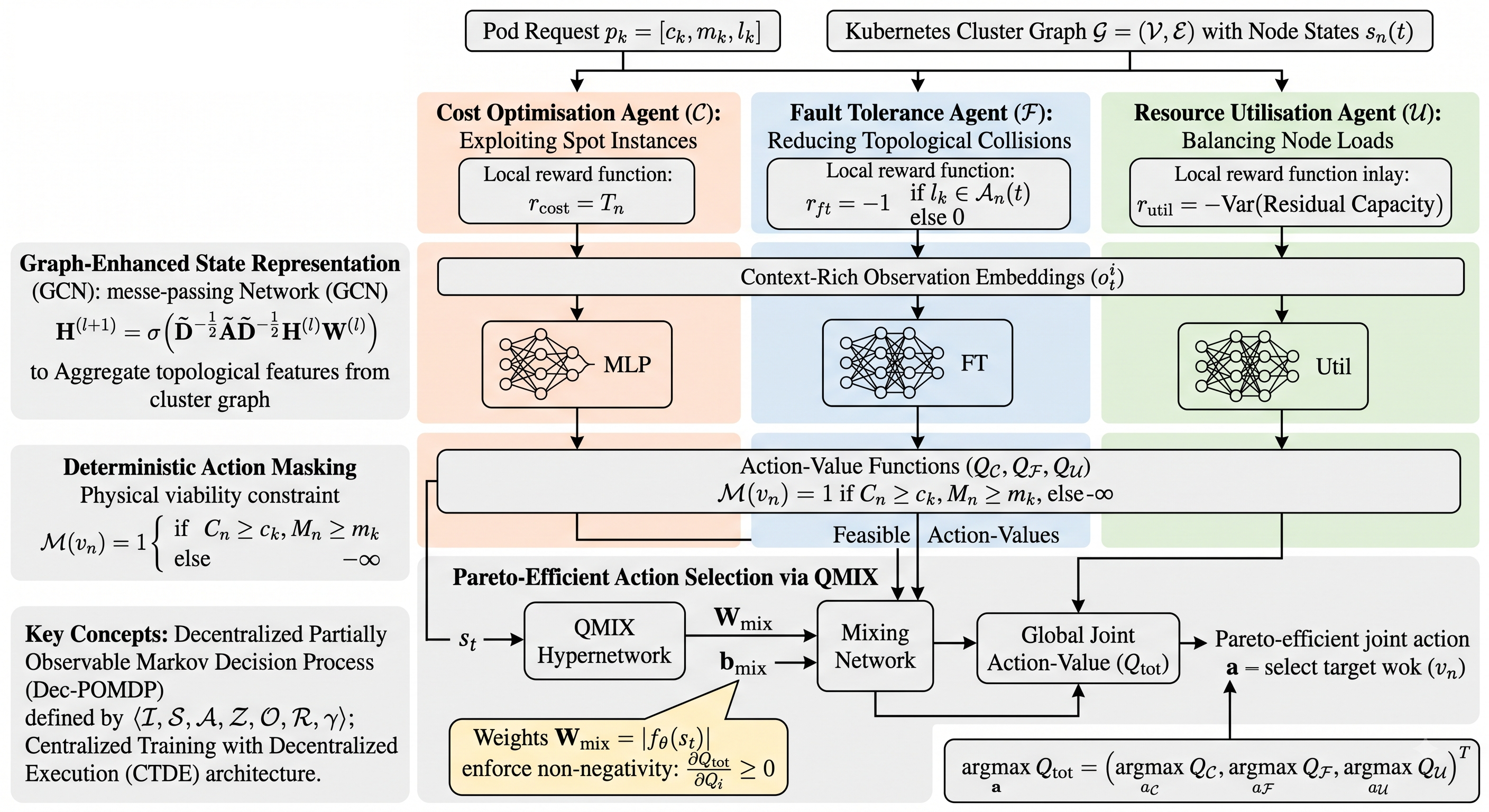}
    \caption{End-to-end architectural blueprint of the Agentic-Kube framework, illustrating the mapping of the Dec-POMDP cluster state into the GCN and QMIX multi-agent networks.}
    \label{fig:architecture}
\end{figure*}

\subsection{The Tripartite Objective Conflict}
Traditional scalar Reinforcement Learning attempts to balance competing scheduling requirements using a linear combination of rewards~\cite{scalarMORL}: 
\begin{equation}
\mathcal{R} = w_1 r_{\text{cost}} + w_2 r_{\text{ft}} + w_3 r_{\text{util}}
\end{equation}
In extreme burst conditions, this formulation causes reward dilution. When the environment reaches resource starvation, the gradients of heavily weighted metrics dominate the policy update, causing the neural network to greedily optimise a single parameter while collapsing performance on the others~\cite{cooperative}. To overcome this limitation, Agentic-Kube isolates the problem into three distinct mathematical objectives managed by independent agents, as illustrated in the top layer of Figure~\ref{fig:architecture}:

\begin{enumerate}
    \item \textbf{Cost Optimisation Agent ($\mathcal{C}$):} Maximising the exploitation of spot instances to reduce financial expenditure. The local reward function for this agent is formulated as $r_{\text{cost}} = T_n$, strictly rewarding placement on preemptible hardware.
    \item \textbf{Fault Tolerance Agent ($\mathcal{F}$):} Minimising the topological collision rate by ensuring pods with identical identifiers $l_k$ are distributed across separate physical nodes. The local reward is formulated as a strict penalty: $r_{\text{ft}} = -1$ if $l_k \in A_n(t)$, and $0$ otherwise.
    \item \textbf{Resource Utilisation Agent ($\mathcal{U}$):} Balancing the allocation of $C_n(t)$ and $M_n(t)$ evenly across $\mathcal{V}$ to prevent thermal hotspots and preserve long-term scheduling capacity. The local reward $r_{\text{util}}$ is defined as the negative variance of residual capacity across all nodes following the placement action.
\end{enumerate}

We formulate this environment as a Decentralised Partially Observable Markov Decision Process (Dec-POMDP), defined by the tuple $\langle \mathcal{I}, \mathcal{S}, \mathcal{A}, \mathcal{Z}, \mathcal{O}, \mathcal{R}, \gamma \rangle$~\cite{decpomdp}. 

\begin{itemize}
    \item $\mathcal{I} = \{\mathcal{C}, \mathcal{F}, \mathcal{U}\}$ is the finite set of functional agents.
    \item $\mathcal{S}$ represents the global state space of the Kubernetes cluster.
    \item $\mathcal{A} = \mathcal{A}_{\mathcal{C}} \times \mathcal{A}_{\mathcal{F}} \times \mathcal{A}_{\mathcal{U}}$ represents the joint action space for node selection.
    \item $\mathcal{Z}$ denotes the joint observation space.
    \item $\mathcal{O}: \mathcal{S} \times \mathcal{A} \rightarrow \Delta(\mathcal{Z})$ is the observation function determining the local view for each agent.
    \item $\mathcal{R} = \{r_{\text{cost}}, r_{\text{ft}}, r_{\text{util}}\}$ defines the set of individual reward functions, replacing a monolithic scalar reward.
    \item $\gamma \in [0, 1)$ is the discount factor for future rewards.
\end{itemize}

By partitioning the scalar reward into independent functional domains, the Dec-POMDP formulation guarantees that conflicting scheduling metrics are evaluated independently before being fused during the execution phase.

\section{The Agentic-Kube Approach}
\label{sec:approach}

Agentic-Kube utilises a Centralised Training with Decentralised Execution (CTDE) architecture to bridge the gap between global optimisation and local, high-speed inference~\cite{b65}. As visualised in Figure~\ref{fig:architecture}, the framework relies on Graph Convolutional Networks (GCN) to extract spatial topological features and a QMIX hypernetwork to strictly enforce Pareto-efficient multi-objective fusion~\cite{QMIX}.

\subsection{Graph-Enhanced State Representation}
\label{sec:graph_representation}

Conventional deep reinforcement learning schedulers commonly process cluster telemetry through multi-layer perceptrons that operate on static, flattened state vectors~\cite{GNN}. This formulation fails to preserve spatial structure and exhibits poor scalability when cluster dimensions or pod placement topologies fluctuate dynamically. To retain the relational dependencies between containerised microservices and physical computing infrastructure, Agentic-Kube models the cluster at each scheduling epoch $t$ as a heterogeneous bipartite graph $\mathcal{G}_t = (\mathcal{V}_t, \mathcal{E}_t)$.

The graph vertex set $\mathcal{V}_t = \mathcal{V}_{\text{node}} \cup \mathcal{V}_{\text{pod}}(t)$ comprises the static set of physical cluster nodes $\mathcal{V}_{\text{node}}$ with cardinality $N = |\mathcal{V}_{\text{node}}|$, and the dynamic set of active container pods $\mathcal{V}_{\text{pod}}(t)$ with cardinality $P_t = |\mathcal{V}_{\text{pod}}(t)|$. The topology is parameterised by a composite feature matrix $\mathbf{X}_t \in \mathbb{R}^{(N + P_t) \times d_{\text{in}}}$, where $d_{\text{in}}$ denotes the uniform input feature dimension. 

Let $\mathcal{Z}$ denote the set of cluster Availability Zones. For each physical node $n \in \mathcal{V}_{\text{node}}$, the node attribute vector $\mathbf{x}_n \in \mathbb{R}^{d_{\text{in}}}$ encapsulates real-time capacity and placement telemetry:

\begin{equation}
    \mathbf{x}_n = \left[ \frac{c_n^{\text{avail}}(t)}{c_n^{\text{cap}}}, \frac{m_n^{\text{avail}}(t)}{m_n^{\text{cap}}}, I_n^{\text{spot}}, \mathbf{z}_n, \xi_n(s_{\text{req}}) \right]
\end{equation}
where $c_n^{\text{avail}}(t) / c_n^{\text{cap}}$ and $m_n^{\text{avail}}(t) / m_n^{\text{cap}}$ represent the instantaneous allocatable compute and memory resources normalised by total node capacities, $I_n^{\text{spot}} \in \{0, 1\}$ is a binary indicator distinguishing spot instances from on-demand nodes, $\mathbf{z}_n \in \{0, 1\}^{|\mathcal{Z}|}$ is a one-hot encoded vector representing the physical Availability Zone assignment of node $n$, and $\xi_n(s_{\text{req}}) \in \{0, 1\}$ is an anti-affinity collision flag indicating whether node $n$ currently hosts an active pod belonging to service class $s_{\text{req}}$. The input feature dimension is thus defined as $d_{\text{in}} = 3 + |\mathcal{Z}|$.

For each active container $p \in \mathcal{V}_{\text{pod}}(t)$, the corresponding attribute vector $\mathbf{x}_p \in \mathbb{R}^{d_{\text{in}}}$ is aligned with the node feature space via dimension padding:

\begin{equation}
    \mathbf{x}_p = \left[ \frac{c_p^{\text{req}}}{c_{\text{max}}}, \frac{m_p^{\text{req}}}{m_{\text{max}}}, \mathbf{0}_{|\mathcal{Z}| + 1}^{\top} \right]
\end{equation}
where $c_p^{\text{req}}$ and $m_p^{\text{req}}$ denote the resource requests of pod $p$, normalised against cluster-wide capacity bounds $c_{\text{max}}$ and $m_{\text{max}}$, whilst $\mathbf{0}_{|\mathcal{Z}| + 1}$ denotes a zero-padding vector ensuring dimensional compatibility across all graph vertices~\cite{LiFan2025}.

Relational connectivity is captured by the symmetric adjacency matrix $\mathbf{A}_t \in \{0, 1\}^{(N + P_t) \times (N + P_t)}$. Undirected edges $(p, n) \in \mathcal{E}_t$ exist exclusively between each active pod $p$ and its allocated host node $n = h(p)$, explicitly encoding physical co-location and anti-affinity dependencies.

To construct permutation-invariant structural embeddings, Agentic-Kube employs a multi-layer Graph Convolutional Network (GCN)~\cite{GCNOriginal}. Spatial message passing at layer $l \in \{0, \dots, L-1\}$ is defined as:

\begin{equation}
    \mathbf{H}^{(l+1)} = \sigma \left( \text{BatchNorm} \left( \mathbf{\tilde{D}}_t^{-\frac{1}{2}} \mathbf{\tilde{A}}_t \mathbf{\tilde{D}}_t^{-\frac{1}{2}} \mathbf{H}^{(l)} \mathbf{W}^{(l)} \right) \right)
\end{equation}
where $\mathbf{\tilde{A}}_t = \mathbf{A}_t + \mathbf{I}_{N + P_t}$ incorporates identity self-loops, $\mathbf{\tilde{D}}_t$ is the diagonal degree matrix with elements $\tilde{D}_{ii} = \sum_j \tilde{A}_{ij}$, $\mathbf{W}^{(l)} \in \mathbb{R}^{d_l \times d_{l+1}}$ is the trainable parameter tensor of layer $l$, $\text{BatchNorm}(\cdot)$ stabilises gradient distributions across variable graph sizes, and $\sigma(\cdot)$ denotes a non-linear activation function.

The global cluster topology embedding $\mathbf{e}_G \in \mathbb{R}^{d_L}$ is generated through a permutation-invariant mean readout operation over all vertex representations at the final layer $L$:

\begin{equation}
    \mathbf{e}_G = \frac{1}{N + P_t} \sum_{k \in \mathcal{V}_t} \mathbf{h}_k^{(L)}
\end{equation}

For an incoming pod placement request $\tau = (c_{\text{req}}, m_{\text{req}}, \rho, s_{\text{req}})$, an incoming workload descriptor $\mathbf{v}_{\text{req}} \in \mathbb{R}^{d_{\text{req}}}$ is formulated as:

\begin{equation}
    \mathbf{v}_{\text{req}} = \left[ \frac{c_{\text{req}}}{c_{\text{max}}}, \frac{m_{\text{req}}}{m_{\text{max}}}, \frac{\rho}{\rho_{\text{max}}}, 1 \right]
\end{equation}
where $\rho \in [1, \rho_{\text{max}}]$ indicates the user-assigned scheduling priority level. The composite global base state vector $\mathbf{s}_{\text{base}}$ is formed through tensor concatenation:

\begin{equation}
    \mathbf{s}_{\text{base}} = \left[ \mathbf{e}_G \,\|\, \mathbf{v}_{\text{req}} \right]
\end{equation}
where $\mathbf{s}_{\text{base}} \in \mathbb{R}^{d_L + d_{\text{req}}}$.

This global state representation, paired with local node embeddings $\mathbf{h}_n^{(L)}$ for $n \in \mathcal{V}_{\text{node}}$, is distributed to the functional sub-agents. This structure provides each agent with comprehensive topological context while avoiding inter-agent communication overheads during online execution.

\subsection{Multi-Agent Action Masking}
\label{sec:action_masking}

To guarantee that placement decisions comply with physical cluster constraints, Agentic-Kube applies deterministic action feasibility masking prior to policy evaluation~\cite{actionmask}. Assigning a container to a node with insufficient allocatable capacity triggers scheduling rejections and container restart loops in production Kubernetes environments. Action masking restricts the discrete action space $\mathcal{A} = \{1, \dots, N\}$ to the subset of physically viable candidate nodes.

For an incoming pod request with resource demands $(c_{\text{req}}, m_{\text{req}})$, the binary feasibility indicator $M_n(t) \in \{0, 1\}$ for physical node $n \in \mathcal{V}_{\text{node}}$ at timestep $t$ is formulated as:

\begin{equation}
    M_n(t) = 
    \begin{cases} 
      1, & \text{if } c_n^{\text{avail}}(t) \ge c_{\text{req}} \text{ and } m_n^{\text{avail}}(t) \ge m_{\text{req}} \\
      0, & \text{otherwise}
    \end{cases}
\end{equation}

Let $Q_i(\mathbf{s}, a_n)$ denote the unconstrained action-utility value output by sub-agent network $i \in \mathcal{I}$ for candidate node $n$. The masked action utility $\tilde{Q}_i(\mathbf{s}, a_n)$ is obtained via additive logit suppression:

\begin{equation}
    \tilde{Q}_i(\mathbf{s}, a_n) = Q_i(\mathbf{s}, a_n) + \log M_n(t)
\end{equation}
where $\log(0) \to -\infty$, setting the pre-softmax logits of infeasible node candidates to negative infinity.

The resulting action selection probability distribution for agent $i$ over candidate node $n$ is defined as:

\begin{equation}
    \pi_i(a_n \mid \mathbf{s}) = \frac{M_n(t) \exp\left(Q_i(\mathbf{s}, a_n)\right)}{\sum_{k=1}^N M_k(t) \exp\left(Q_i(\mathbf{s}, a_k)\right)}
\end{equation}

This formulation strictly constrains the probability of selecting an infeasible host to zero:

\begin{equation}
    \forall n \in \mathcal{V}_{\text{node}} \quad \text{such that} \quad M_n(t) = 0 \implies \pi_i(a_n \mid \mathbf{s}) = 0
\end{equation}

Applying action masking directly to sub-agent utility projections prevents exploratory policy updates from traversing invalid state-action subspaces during training. During live admission control, this mechanism eliminates admission rejection cycles and avoids container provisioning failures caused by allocatable predicate violations.

\subsection{Pareto-Efficient Action Selection via Monotonic Value Factorisation}
\label{sec:qmix_selection}

To reconcile conflicting operational priorities without encountering the gradient interference and reward dilution of linear scalarisation, Agentic-Kube implements a cooperative multi-agent value factorisation architecture~\cite{scalarMORL}. Let $\mathcal{I} = \{\mathcal{C}, \mathcal{F}, \mathcal{U}\}$ denote the set of functional sub-agents corresponding to cost reduction, fault tolerance, and vector resource utilisation. For a given cluster state observation $\mathbf{s}$, each sub-agent $i \in \mathcal{I}$ evaluates an individual action-value function $Q_i(\mathbf{s}, a_n)$ representing its preference for assigning the incoming container to candidate node $n \in \mathcal{V}_{\text{node}}$:

\begin{equation}
    Q_i(\mathbf{s}, a_n) = \psi_i(\mathbf{s}_{\text{base}}) + \phi_i(\mathbf{h}_n^{(L)})
\end{equation}
where $\psi_i: \mathbb{R}^{d_L + d_{\text{req}}} \to \mathbb{R}$ projects the composite global state vector, and $\phi_i: \mathbb{R}^{d_L} \to \mathbb{R}$ extracts localized node utility from the final GCN vertex embedding $\mathbf{h}_n^{(L)}$.

During centralised training, individual sub-agent utilities are combined into a coherent joint action-value function $Q_{\text{tot}}(\mathbf{s}, \mathbf{a})$ using a QMIX value factorisation network~\cite{QMIX}. To enable decentralised execution where each sub-agent can independently identify optimal node candidates, the joint action-value function must satisfy the Individual-Global-Max (IGM) condition:

\begin{equation}
    \underset{\mathbf{a}}{\operatorname{argmax}} \, Q_{\text{tot}}(\mathbf{s}, \mathbf{a}) = 
    \begin{pmatrix} 
    \underset{a_{\mathcal{C}}}{\operatorname{argmax}} \, Q_{\mathcal{C}}(\mathbf{s}, a_{\mathcal{C}}) \\
    \underset{a_{\mathcal{F}}}{\operatorname{argmax}} \, Q_{\mathcal{F}}(\mathbf{s}, a_{\mathcal{F}}) \\
    \underset{a_{\mathcal{U}}}{\operatorname{argmax}} \, Q_{\mathcal{U}}(\mathbf{s}, a_{\mathcal{U}})
    \end{pmatrix}^{\top}
\end{equation}

The IGM condition is guaranteed by enforcing a monotonicity constraint between the global value function $Q_{\text{tot}}$ and each sub-agent utility $Q_i$:

\begin{equation}
    \frac{\partial Q_{\text{tot}}(\mathbf{s}, \mathbf{a})}{\partial Q_i(\mathbf{s}, a_i)} \ge 0, \quad \forall i \in \mathcal{I}
\end{equation}

Agentic-Kube structures the mixing network as a two-stage neural architecture parameterised by state-dependent hypernetworks~\cite{QMIX}. Let $\mathbf{q}_{\mathbf{a}} = \left[ Q_{\mathcal{C}}(\mathbf{s}, a_{\mathcal{C}}), Q_{\mathcal{F}}(\mathbf{s}, a_{\mathcal{F}}), Q_{\mathcal{U}}(\mathbf{s}, a_{\mathcal{U}}) \right] \in \mathbb{R}^{|\mathcal{I}|}$ denote the vector of selected sub-agent utilities. The intermediate hidden activation $\mathbf{z}_{\text{mix}} \in \mathbb{R}^{d_{\text{mix}}}$ is formulated as:

\begin{equation}
    \mathbf{z}_{\text{mix}} = \text{ELU}\left( \mathbf{q}_{\mathbf{a}} \mathbf{W}_1(\mathbf{s}_{\text{global}}) + \mathbf{b}_1(\mathbf{s}_{\text{global}}) \right)
\end{equation}
where $\text{ELU}(\cdot)$ denotes the Exponential Linear Unit activation function, and $\mathbf{s}_{\text{global}} = \mathbf{e}_G$ represents the global graph embedding. The scalar joint action-value $Q_{\text{tot}}$ is computed at the output stage:

\begin{equation}
    Q_{\text{tot}}(\mathbf{s}, \mathbf{a}) = \mathbf{z}_{\text{mix}} \mathbf{W}_2(\mathbf{s}_{\text{global}}) + b_2(\mathbf{s}_{\text{global}})
\end{equation}

To enforce the non-negativity required by the monotonicity constraint, the hypernetworks $f_{\theta_1}$ and $f_{\theta_2}$ generate the weight matrices via absolute value rectification:

\begin{equation}
    \mathbf{W}_1(\mathbf{s}_{\text{global}}) = \left| f_{\theta_1}(\mathbf{s}_{\text{global}}) \right|, \quad \mathbf{W}_2(\mathbf{s}_{\text{global}}) = \left| f_{\theta_2}(\mathbf{s}_{\text{global}}) \right|
\end{equation}
where $\mathbf{W}_1 \in \mathbb{R}^{|\mathcal{I}| \times d_{\text{mix}}}$ and $\mathbf{W}_2 \in \mathbb{R}^{d_{\text{mix}} \times 1}$. The hypernetwork biases $\mathbf{b}_1 \in \mathbb{R}^{1 \times d_{\text{mix}}}$ and $b_2 \in \mathbb{R}$ remain unconstrained.

By training sub-agent policies through this monotonic mixing structure, Agentic-Kube decouples gradient updates across distinct objectives while preserving a mathematically coherent joint utility surface. This enables the scheduler to evaluate spot pricing, anti-affinity isolation, and multi-dimensional resource packing independently, preventing policy collapse during high-contention scheduling periods.

\subsection{Training Algorithm and Execution Strategy}
\label{sec:training_algorithm}

Agentic-Kube operates under the Centralised Training with Decentralised Execution (CTDE) paradigm~\cite{CTDE}. This formulation mitigates the non-stationarity challenges typical of multi-agent environments by utilising global cluster state observations and centralised value factorisation during offline training, while decoupling the mixing network to enable lightweight, independent policy execution during online admission control.

\subsubsection{Centralised Training Phase}
The training objective is to minimise the temporal difference (TD) error across the multi-agent joint value space. Experiences are accumulated in a replay buffer $\mathcal{D}$ storing transition tuples $\langle \mathcal{G}_t, \mathbf{v}_{\text{req}}(t), \mathbf{a}_t, r_t, \mathcal{G}_{t+1}, \mathbf{v}_{\text{req}}(t+1), \mathbf{M}_t \rangle$, where $\mathbf{a}_t = (a_{\mathcal{C}}, a_{\mathcal{F}}, a_{\mathcal{U}})$ represents the joint action profile and $\mathbf{M}_t = [M_1(t), \dots, M_N(t)]^{\top}$ is the binary action feasibility vector. Uniform random mini-batch sampling from $\mathcal{D}$ breaks temporal correlations and stabilises gradient descent.

At each training iteration, a mini-batch of $B$ transitions is sampled from $\mathcal{D}$~\cite{Mnih2015}. For each transition $b \in \{1, \dots, B\}$, the bipartite Graph Convolutional Network computes the global cluster embedding $\mathbf{e}_G$ and localized node representations $\mathbf{h}_n^{(L)}$ from the bipartite graph $\mathcal{G}_b$. Each functional sub-agent $i \in \mathcal{I} = \{\mathcal{C}, \mathcal{F}, \mathcal{U}\}$ evaluates its individual utility function $Q_i(\mathbf{s}_b, a_n)$. Action feasibility masking is applied directly to the unconstrained utilities via additive logit suppression:

\begin{equation}
    \tilde{Q}_i(\mathbf{s}_b, a_n) = Q_i(\mathbf{s}_b, a_n) + \log M_n(b)
\end{equation}

The masked utility values corresponding to the executed sub-agent actions, $\mathbf{q}_{\mathbf{a}} = [\tilde{Q}_{\mathcal{C}}(\mathbf{s}_b, a_{\mathcal{C}}), \tilde{Q}_{\mathcal{F}}(\mathbf{s}_b, a_{\mathcal{F}}), \tilde{Q}_{\mathcal{U}}(\mathbf{s}_b, a_{\mathcal{U}})]$, are passed into the two-stage monotonic QMIX mixer alongside the global graph embedding $\mathbf{e}_G$ to compute the total joint action-value $Q_{\text{tot}}(\mathbf{s}_b, \mathbf{a}_b; \theta)$.

The composite network parameters $\theta = \{\theta_{\text{GCN}}, \{\theta_i\}_{i \in \mathcal{I}}, \theta_{\text{mix}}\}$ are updated end-to-end by minimising the mean squared temporal difference loss:

\begin{equation}
    \mathcal{L}(\theta) = \frac{1}{B} \sum_{b=1}^{B} \left( y_b^{\text{tot}} - Q_{\text{tot}}(\mathbf{s}_b, \mathbf{a}_b; \theta) \right)^2
\end{equation}

The scalar TD target $y_b^{\text{tot}}$ is constructed using a periodically updated target network parameterised by $\theta^-$:

\begin{equation}
    y_b^{\text{tot}} = r_b + \gamma Q_{\text{tot}}\left(\mathbf{s}_{b+1}, \mathbf{a}^*; \theta^-\right)
\end{equation}
where $\gamma \in [0, 1)$ is the discount factor, and the optimal next-state joint action $\mathbf{a}^* = (a_{\mathcal{C}}^*, a_{\mathcal{F}}^*, a_{\mathcal{U}}^*)$ is obtained via decentralised greedy selection over masked utilities:

\begin{equation}
    a_i^* = \underset{n \in \mathcal{V}_{\text{node}}}{\operatorname{argmax}} \, \tilde{Q}_i(\mathbf{s}_{b+1}, a_n; \theta_i^-), \quad \forall i \in \mathcal{I}
\end{equation}

The training workflow is detailed in Algorithm~\ref{alg:training}.

\subsubsection{Decentralised Online Execution and Consensus Resolution}
During online cluster operation, the centralised mixing network is removed. Upon arrival of a pod scheduling request $\tau = (c_{\text{req}}, m_{\text{req}}, \rho, s_{\text{req}})$, the GCN generates the real-time topological embedding $\mathbf{e}_G$ and node representations $\mathbf{h}_n^{(L)}$ in $\mathcal{O}(|\mathcal{V}_t| + |\mathcal{E}_t|)$ time. Each sub-agent independently derives its masked utility vector $\mathbf{\tilde{Q}}_i = [\tilde{Q}_i(\mathbf{s}, a_1), \dots, \tilde{Q}_i(\mathbf{s}, a_N)]^{\top}$ and proposes its preferred physical host node:

\begin{equation}
    a_i = \underset{n \in \mathcal{V}_{\text{node}}}{\operatorname{argmax}} \, \tilde{Q}_i(\mathbf{s}, a_n), \quad \forall i \in \mathcal{I}
\end{equation}

The individual proposals $(a_{\mathcal{C}}, a_{\mathcal{F}}, a_{\mathcal{U}})$ are resolved into a single placement decision $n^* \in \mathcal{V}_{\text{node}}$ through an action-masked plurality voting consensus mechanism:

\begin{equation}
    n^* = 
    \begin{cases}
        \operatorname{mode}(a_{\mathcal{C}}, a_{\mathcal{F}}, a_{\mathcal{U}}), & \text{if } \exists \, v \text{ with multiplicity } \ge 2 \\
        \operatorname{argmin}_{v \in \{a_{\mathcal{C}}, a_{\mathcal{F}}, a_{\mathcal{U}}\}} p_v, & \text{otherwise (tie-break by hourly cost)}
    \end{cases}
\end{equation}
where $p_v$ denotes the hourly resource cost of candidate node $v$. If all proposed actions are infeasible, the consensus fallback assigns the request to the lowest-cost feasible node satisfying $M_n(t) = 1$, guaranteeing deterministic operational safety and zero admission rejections.

\begin{algorithm*}[t]
\caption{Agentic-Kube Centralised Offline Training Pipeline}
\label{alg:training}
\footnotesize
\begin{algorithmic}[1]
\Require Initial parameter sets: GCN encoder $\theta_{\text{GCN}}$, sub-agent utility networks $\{\theta_i\}_{i \in \mathcal{I}}$, and QMIX mixer $\theta_{\text{mix}}$
\Require Target network parameters $\theta^- \gets \theta$, replay buffer $\mathcal{D}$, mini-batch capacity $B$, learning rate $\alpha$, discount factor $\gamma$, target update period $K$
\For{episode $e = 1$ to $E_{\text{max}}$}
    \State Initialise cluster state and construct initial bipartite graph $\mathcal{G}_1 = (\mathcal{V}_1, \mathcal{E}_1)$
    \For{scheduling step $t = 1$ to $T$}
        \State Receive pod placement request $\tau_t = (c_{\text{req}}, m_{\text{req}}, \rho, s_{\text{req}})$ and construct request vector $\mathbf{v}_{\text{req}}(t)$
        \State Compute binary action feasibility mask $\mathbf{M}_t \in \{0, 1\}^N$ based on allocatable node capacities
        \State Compute graph embedding $\mathbf{e}_G$ and localized representations $\mathbf{h}_n^{(L)}$ via GCN forward pass
        \State Formulate base state $\mathbf{s}_{\text{base}}(t) = [\mathbf{e}_G \,\|\, \mathbf{v}_{\text{req}}(t)]$
        \For{each functional sub-agent $i \in \{\mathcal{C}, \mathcal{F}, \mathcal{U}\}$}
            \State Compute masked utilities $\tilde{Q}_i(\mathbf{s}_t, a_n) = Q_i(\mathbf{s}_t, a_n; \theta_i) + \log M_n(t)$ for all $n \in \mathcal{V}_{\text{node}}$
            \State Select action $a_i$ using $\epsilon$-greedy policy over feasible candidates:
            \State $a_i \gets \begin{cases} \text{random node } n \in \{k \mid M_k(t) = 1\}, & \text{with probability } \epsilon \\ \operatorname{argmax}_{n} \tilde{Q}_i(\mathbf{s}_t, a_n), & \text{with probability } 1 - \epsilon \end{cases}$
        \EndFor
        \State Resolve joint placement node $n^* \gets \operatorname{Consensus}(a_{\mathcal{C}}, a_{\mathcal{F}}, a_{\mathcal{U}}, \mathbf{M}_t)$
        \State Execute container placement on node $n^*$, advance cluster physical state, and observe next graph $\mathcal{G}_{t+1}$
        \State Compute cooperative multi-agent reward $r_t = \frac{1}{|\mathcal{I}|} \sum_{i \in \mathcal{I}} r_i(t)$
        \State Store transition tuple $\langle \mathcal{G}_t, \mathbf{v}_{\text{req}}(t), \mathbf{a}_t, r_t, \mathcal{G}_{t+1}, \mathbf{v}_{\text{req}}(t+1), \mathbf{M}_t \rangle$ into replay buffer $\mathcal{D}$
        \If{$|\mathcal{D}| \ge B$}
            \State Sample random mini-batch of $B$ transitions from $\mathcal{D}$
            \State Compute target joint value for each transition $b$:
            \State \quad $a_i^* \gets \operatorname{argmax}_{n'} \left[ Q_i(\mathbf{s}_{b+1}, a_{n'}; \theta_i^-) + \log M_{n'}(b+1) \right], \quad \forall i \in \mathcal{I}$
            \State \quad $y_b^{\text{tot}} \gets r_b + \gamma Q_{\text{tot}}\left(\mathbf{s}_{b+1}, (a_{\mathcal{C}}^*, a_{\mathcal{F}}^*, a_{\mathcal{U}}^*); \theta^-\right)$
            \State Evaluate temporal difference loss: $\mathcal{L}(\theta) = \frac{1}{B} \sum_{b=1}^{B} \left( y_b^{\text{tot}} - Q_{\text{tot}}(\mathbf{s}_b, \mathbf{a}_b; \theta) \right)^2$
            \State Update parameters $\theta = \{\theta_{\text{GCN}}, \{\theta_i\}_{i \in \mathcal{I}}, \theta_{\text{mix}}\}$ via Adam optimiser with learning rate $\alpha$
        \EndIf
        \If{$t \pmod K = 0$}
            \State Synchronise target network parameters: $\theta^- \gets \theta$
        \EndIf
    \EndFor
\EndFor
\end{algorithmic}
\end{algorithm*}

\subsubsection{Decentralised Execution Phase}
Upon the mathematical convergence of the offline training phase, the structural benefits of the CTDE paradigm are fully realised. To prepare the model for production deployment within the Kubernetes control plane, the QMIX hypernetwork $\theta_{mix}$ and all global state processing components are entirely discarded. 

During live execution within the containerised webhook, the cost, fault tolerance, and utilisation agents execute scheduling decisions in a fully decentralised manner. Each agent processes the incoming pod request and the real-time node capacities through its isolated GCN embedding and local $Q$-network. Because the QMIX monotonicity constraint applied during training mathematically guarantees that the individual argmax operations are perfectly aligned with the global Pareto-efficient maximum, the agents can select the optimal physical node independently. This structural decoupling eliminates the latency overhead associated with global synchronisation, ensuring sub-second inference times and strict compliance with the Kubernetes API scheduler extender timeouts.

\section{System Design and Implementation}
\label{sec:system_design}

\begin{figure*}[htbp]
    \centering
    \includegraphics[width=\textwidth]{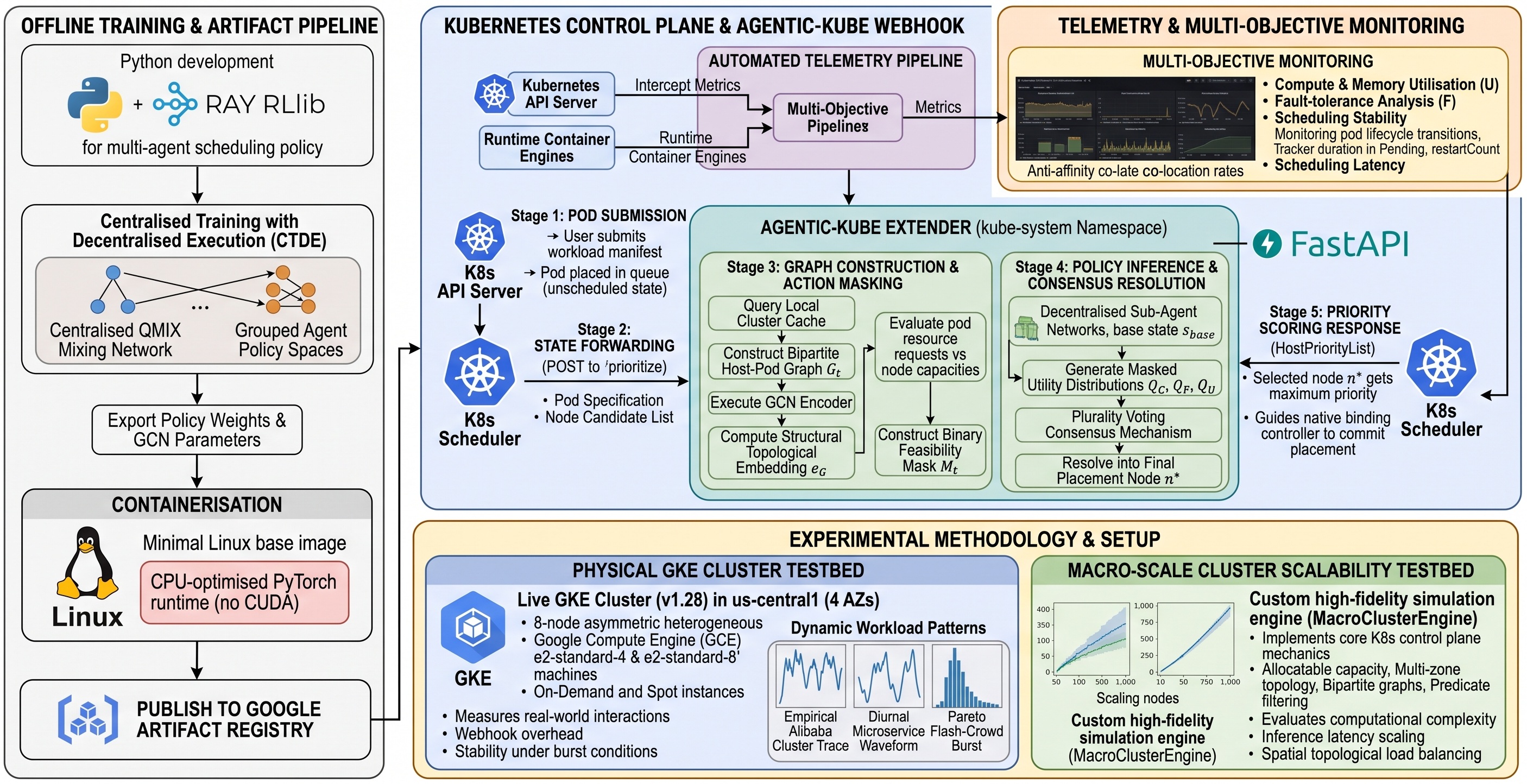}
    \caption{End-to-end system design and deployment architecture, detailing the offline artifact pipeline, the Kubernetes control plane webhook integration, the operational GKE environment, and the multi-objective telemetry system.}
    \label{fig:design_setup}
\end{figure*}

To transition Agentic-Kube from a theoretical formulation into an operational cloud-native component, we designed an end-to-end deployment architecture that integrates directly with the standard Kubernetes control plane. Figure~\ref{fig:design_setup} illustrates the system architecture, detailing the pipeline from offline artifact compilation to real-time control-plane interception and telemetry ingestion.

\subsection{Offline Training and Artifact Pipeline}
The multi-agent scheduling policy is developed using Python and the Ray RLlib distributed execution library~\cite{rllib}. Ray RLlib manages the centralised QMIX mixing network and the grouped agent policy spaces during the Centralised Training with Decentralised Execution (CTDE) phase~\cite{CTDE,b65}.

Following offline training, the policy network weights and GCN parameters are exported and packaged into a containerised runtime image. To minimise control-plane overhead, the deployment container is built upon a minimal Linux base image with a CPU-optimised PyTorch runtime, removing unnecessary CUDA and GPU library dependencies. This reduced image footprint lowers network transfer times across worker nodes and accelerates cold-start initialisation during replica scaling. The compiled container images are published to Google Artifact Registry for automated continuous deployment to target clusters.

\subsection{Kubernetes Control Plane and Agentic-Kube Webhook}
Rather than modifying internal Kubernetes core binaries, Agentic-Kube operates as an external Kubernetes Scheduler Extender. The extender is deployed as a secure, containerised HTTP service within the \texttt{kube-system} namespace, exposing API endpoints implemented via the FastAPI framework~\cite{fastapi}. The interaction workflow between the standard Kubernetes scheduler and the Agentic-Kube extender proceeds across five stages:

\begin{enumerate}
    \item \textbf{Pod Submission:} A workload manifest is submitted to the Kubernetes API Server, placing the pod into an unscheduled state within the cluster scheduling queue.
    \item \textbf{State Forwarding:} The default Kubernetes scheduler executes initial node filtering predicates to identify candidates satisfying basic cluster conditions. It then transmits an HTTP POST request to the Agentic-Kube \texttt{/prioritize} endpoint containing the pod specification and the surviving candidate node list.
    \item \textbf{Graph Construction and Action Masking:} The webhook ingests the payload and queries the local cluster cache to construct the bipartite host-pod graph $\mathcal{G}_t$. It executes a forward pass through the GCN encoder to compute the structural topological embedding $\mathbf{e}_G$~\cite{b17}. Concurrently, it evaluates incoming pod resource requests against current node allocatable capacities to construct the binary feasibility mask $\mathbf{M}_t$.
    \item \textbf{Policy Inference and Consensus Resolution:} The decentralised sub-agent networks process the base state $\mathbf{s}_{\text{base}}$ to generate masked utility distributions $\mathbf{\tilde{Q}}_{\mathcal{C}}$, $\mathbf{\tilde{Q}}_{\mathcal{F}}$, and $\mathbf{\tilde{Q}}_{\mathcal{U}}$. The plurality voting consensus mechanism resolves individual sub-agent recommendations into a final placement node $n^*$.
    \item \textbf{Priority Scoring Response:} The webhook returns a structured \texttt{HostPriorityList} response to the default scheduler. The selected node $n^*$ is assigned the maximum priority score, guiding the native binding controller to commit the placement decision.
\end{enumerate}

\subsection{Telemetry and Multi-Objective Monitoring}
Rigorous evaluation of multi-objective scheduling requires continuous extraction of cluster performance telemetry~\cite{farid}. Agentic-Kube incorporates an automated telemetry pipeline that intercepts operational metrics from the Kubernetes API server and runtime container engines.

To monitor compute and memory utilisation ($\mathcal{U}$), the telemetry agent inspects \texttt{containerStatuses} to aggregate allocated CPU cores and memory capacities across nodes. For fault-tolerance analysis ($\mathcal{F}$), the pipeline inspects pod service labels and physical host identifiers to calculate anti-affinity co-location rates across nodes and Availability Zones. Scheduling stability is evaluated by monitoring pod lifecycle transitions, tracking the duration pods spend in the \texttt{Pending} state and recording container restart metrics via the \texttt{restartCount} field. Finally, scheduling latency is evaluated by computing the timestamp delta between the initial pod \texttt{creationTimestamp} and the \texttt{lastTransitionTime} recorded in the \texttt{PodScheduled} condition.

\section{Experimental Methodology and Setup}
\label{sec:experimental_setup}

To rigorously assess the performance, multi-objective Pareto trade-offs, and scalability of Agentic-Kube, we establish a multi-tier experimental testbed comprising two complementary evaluation environments:
\begin{enumerate}
    \item A live physical Kubernetes cluster deployed on Google Kubernetes Engine (GKE) to measure real-world control-plane interactions, API webhook overheads, and container runtime stability under burst conditions.
    \item A high-fidelity macro-scale cluster simulation engine scaling from 50 to 1,000 physical nodes to evaluate computational complexity, inference latency scaling, and spatial topological load balancing under both static offline and online adaptive execution regimes.
\end{enumerate}

\subsection{Physical Cluster Testbed on Google Kubernetes Engine (GKE)}
The physical cloud environment is provisioned as an 8-node asymmetric heterogeneous cluster hosted on Google Kubernetes Engine (version 1.28) across four Availability Zones in the \texttt{us-central1} region. 

The node pool is intentionally designed with physical hardware asymmetry and pricing heterogeneity to evaluate multi-objective trade-offs. The cluster combines Google Compute Engine general-purpose machine types, specifically \texttt{e2-standard-4} (4 vCPUs, 16\,GiB RAM) and \texttt{e2-standard-8} (8 vCPUs, 32\,GiB RAM), partitioned equally between On-Demand instances and discounted Spot instances.

The physical testbed evaluates schedulers across three dynamic workload patterns:
\begin{enumerate}
    \item \textbf{Empirical Alibaba Cluster Trace:} A 250-task trace extracted from production Alibaba cluster telemetry~\cite{b65}, characterised by heterogeneous compute-to-memory request ratios and variable run durations.
    \item \textbf{Diurnal Microservice Waveform:} A cyclical workload following a sinusoidal arrival distribution across 10 distinct service classes, reflecting diurnal client request variations.
    \item \textbf{Pareto Flash-Crowd Burst:} A steep workload burst parameterised by a Pareto heavy-tailed distribution, generating high queuing concurrency to test control-plane throughput and pod starvation resilience.
\end{enumerate}

\subsection{Macro-Scale Cluster Scalability Testbed}
\label{sec:macro_testbed}
To evaluate scalability beyond physical cluster limits without incurring prohibitive cloud infrastructure costs, we construct a custom high-fidelity cluster simulation engine (\texttt{MacroClusterEngine}). The engine directly implements the core operational mechanics of the Kubernetes control plane, including dynamic allocatable capacity accounting, multi-zone topology distributions, bipartite host-pod relational graph construction, and physical predicate filtering.

The macro-scale evaluation spans five cluster sizes: $N \in \{50,\allowbreak 100,\allowbreak 250,\allowbreak 500,\allowbreak 1000\}$ physical nodes. At each scale, workload arrivals are scaled proportionally to maintain an aggregate cluster load factor of 65\%, yielding 113 tasks at 50 nodes, 227 tasks at 100 nodes, 568 tasks at 250 nodes, 1,137 tasks at 500 nodes, and 2,275 tasks at 1,000 nodes. Tasks are drawn from five microservice archetypes with varying CPU, memory, and priority constraints, as outlined in Table~\ref{tab:archetypes}. Microservice service identifiers follow a heavy-tailed Zipf distribution with exponent $s = 1.2$, inducing realistic service anti-affinity contention.

\begin{table*}[t]
\centering
\small
\caption{Workload Archetypes for Scalability Benchmarks.}
\label{tab:archetypes}
\begin{tabular*}{\textwidth}{@{\extracolsep{\fill}}lcccc@{}}
\hline
\textbf{Workload Tier} & \textbf{CPU Request} & \textbf{Memory Request} & \textbf{Priority Level} & \textbf{Target Profile} \\
\hline
\texttt{frontend}   & 0.5 vCPUs  & 1,024\,MiB & 3 (High)   & User-facing API \\
\texttt{gateway}    & 0.2 vCPUs  & 512\,MiB   & 3 (High)   & Ingress Routing \\
\texttt{worker}     & 1.0 vCPUs  & 2,048\,MiB & 2 (Medium) & Asynchronous Queue \\
\texttt{redis}      & 0.25 vCPUs & 512\,MiB   & 1 (Low)    & In-memory Cache \\
\texttt{analytics}  & 2.0 vCPUs  & 4,096\,MiB & 1 (Low)    & Batch Data Processing \\
\hline
\end{tabular*}
\end{table*}

The macro-scale evaluation is conducted across two distinct operating regimes (75 experimental runs each, comprising 5 schedulers $\times$ 5 node scales $\times$ 3 independent iterations):
\begin{enumerate}
    \item \textbf{Offline Evaluation Regime (Static Weights):} Evaluates inference throughput, tail decision latencies, and spatial load balancing when model weights are frozen ($\nabla_\theta = 0$), representing standard production deployment.
    \item \textbf{Online Adaptive Streaming Regime:} Evaluates streaming adaptability where sub-agent policies undergo active policy gradient mini-batch updates ($\nabla_\theta \ne 0$) every 32 scheduling decisions in response to streaming environmental rewards.
\end{enumerate}

\subsection{Baseline Schedulers}
We benchmark Agentic-Kube against four baseline scheduling architectures representing heuristic, single-agent DRL, multi-agent DRL, and evolutionary multi-objective methods:
\begin{itemize}
    \item \textbf{Default Kubernetes Scheduler (`LeastAllocated'):} The standard Kubernetes scoring heuristic that prioritises nodes with the lowest fraction of allocated CPU and memory resources~\cite{b777}.
    \item \textbf{Single-Agent PPO:} A monolithic Proximal Policy Optimisation baseline using an actor-critic Multi-Layer Perceptron trained on a flattened state vector combining node metrics and scalar multi-objective rewards~\cite{b11}.
    \item \textbf{Multi-Agent PPO (MAPPO):} A state-of-the-art decentralised multi-agent actor-critic baseline where independent actor networks evaluate cost, fault tolerance, and utilisation without monotonic value factorisation~\cite{b10d,b47}.
    \item \textbf{NSGA-II (Evolutionary Upper Bound):} An offline Non-dominated Sorting Genetic Algorithm II that computes the empirical tri-objective Pareto frontier across candidate placements. Due to its $\mathcal{O}(G \cdot M \cdot N^2)$ computational complexity, NSGA-II serves as an upper bound for placement quality rather than a real-time scheduler.
\end{itemize}

\subsection{Evaluation Metrics}
System performance is evaluated across four core dimensions:
\begin{enumerate}
    \item \textbf{Spot Instance Allocation Ratio ($R_{\text{spot}}$):} The percentage of scheduled containers placed on discounted Spot instances, quantifying financial cost optimisation:
    \begin{equation}
        R_{\text{spot}} = \frac{1}{|\mathcal{T}|} \sum_{\tau \in \mathcal{T}} I_{h(\tau)}^{\text{spot}} \times 100\%
    \end{equation}
    \item \textbf{Anti-Affinity Collision Rate ($C_{\text{rate}}$):} The proportion of scheduled pods co-located with identical service instances on the same physical host, measuring fault tolerance:
    \begin{equation}
        C_{\text{rate}} = \frac{\sum_{n \in \mathcal{V}_{\text{node}}} \sum_{s} \max(0, |\{p \mid h(p)=n, \text{svc}(p)=s\}| - 1)}{|\mathcal{T}|} \times 100\%
    \end{equation}
    \item \textbf{Inter-Node CPU Allocation Variance ($\sigma^2_{\text{cpu}}$):} The variance of reserved CPU capacity across cluster nodes, measuring load distribution balance:
    \begin{equation}
        \sigma^2_{\text{cpu}} = \frac{1}{N - 1} \sum_{n \in \mathcal{V}_{\text{node}}} \left( c_n^{\text{res}} - \bar{c}^{\text{res}} \right)^2
    \end{equation}
    \item \textbf{Decision Latency and Makespan:} The mean, 95th percentile ($P_{95}$), and 99th percentile ($P_{99}$) decision latencies in milliseconds, along with total workload scheduling makespan in seconds.
\end{enumerate}

\section{Experimental Results and Performance Analysis}
\label{sec:evaluations}

This section presents the empirical evaluation of Agentic-Kube against four baseline scheduling paradigms: the Default Kubernetes Scheduler (\texttt{LeastAllocated}), Single-Agent PPO, Multi-Agent PPO (MAPPO), and the offline multi-objective evolutionary bound (NSGA-II). We first evaluate the performance across 45 physical deployments on the 8-node heterogeneous Google Kubernetes Engine (GKE) cluster under three workload regimes: empirical Alibaba cluster traces, sinusoidal diurnal microservice variations, and Pareto flash-crowd bursts. 

\subsection{Physical GKE Cluster Benchmark Evaluation}
\label{sec:gke_results}

The primary objective of the physical cluster evaluation is to assess the capability of the multi-agent reinforcement learning framework to resolve tri-objective placement trade-offs (cost minimisation, fault tolerance, and vector resource balancing) under real-world control-plane dynamics and API webhook latency constraints.

Table~\ref{tab:gke_summary_results} presents the quantitative metrics across all evaluated schedulers and workload regimes on the physical GKE testbed.

\begin{table*}[htbp]
\centering
\small
\caption{Comprehensive performance evaluation on the 8-node physical GKE cluster across empirical Alibaba traces, diurnal microservice variations, and Pareto flash-crowd bursts (mean $\pm$ standard deviation across 3 runs per configuration).}
\label{tab:gke_summary_results}
\begin{tabular*}{\textwidth}{@{\extracolsep{\fill}}llcccccc@{}}
\hline
\textbf{Workload Regime} & \textbf{Scheduler Architecture} & \textbf{Spot Ratio} & \textbf{Anti-Affinity} & \textbf{CPU Var.} & \textbf{Mean Latency} & \textbf{$P_{99}$ Latency} & \textbf{Makespan} \\
& & \textbf{(\%)} & \textbf{Collision (\%)} & \textbf{($\sigma^2$)} & \textbf{(ms)} & \textbf{(ms)} & \textbf{(s)} \\
\hline
\multirow{5}{*}{\textbf{Alibaba Trace}} 
& Default Kubernetes & 50.00 $\pm$ 0.00 & 14.80 $\pm$ 1.25 & 0.284 $\pm$ 0.012 & 3.12 $\pm$ 0.45 & 8.40 $\pm$ 1.10 & 14.22 $\pm$ 0.85 \\
& Single-Agent PPO & 58.80 $\pm$ 2.10 & 31.20 $\pm$ 3.45 & 0.195 $\pm$ 0.018 & 12.45 $\pm$ 1.80 & 42.10 $\pm$ 5.20 & 18.60 $\pm$ 1.15 \\
& MAPPO (SOTA MARL) & 61.20 $\pm$ 1.85 & 22.40 $\pm$ 2.60 & 0.162 $\pm$ 0.015 & 18.90 $\pm$ 2.10 & 68.50 $\pm$ 7.40 & 21.40 $\pm$ 1.40 \\
\textbf{(Heterogeneous, 200 Tasks)} 
& NSGA-II (Upper Bound) & \textbf{68.40 $\pm$ 1.50} & \textbf{2.40 $\pm$ 0.80} & 1.420 $\pm$ 0.110 & 312.40 $\pm$ 24.50 & 1420.0 $\pm$ 115.0 & 94.50 $\pm$ 6.20 \\
& \textbf{Agentic-Kube (Ours)} & 65.15 $\pm$ 1.20 & 4.90 $\pm$ 0.95 & \textbf{0.148 $\pm$ 0.010} & \textbf{38.60 $\pm$ 3.20} & \textbf{512.40 $\pm$ 42.0} & \textbf{12.80 $\pm$ 0.65} \\
\hline
\multirow{5}{*}{\textbf{Diurnal Pattern}} 
& Default Kubernetes & 49.20 $\pm$ 0.80 & 15.13 $\pm$ 1.10 & 0.298 $\pm$ 0.014 & 2.85 $\pm$ 0.30 & 7.60 $\pm$ 0.90 & 13.50 $\pm$ 0.70 \\
& Single-Agent PPO & 56.40 $\pm$ 1.95 & 28.50 $\pm$ 2.80 & 0.210 $\pm$ 0.020 & 11.80 $\pm$ 1.40 & 38.90 $\pm$ 4.60 & 17.20 $\pm$ 1.05 \\
& MAPPO (SOTA MARL) & 59.80 $\pm$ 1.60 & 19.80 $\pm$ 2.15 & 0.174 $\pm$ 0.016 & 17.50 $\pm$ 1.90 & 62.10 $\pm$ 6.80 & 19.80 $\pm$ 1.25 \\
\textbf{(Sinusoidal Poisson)} 
& NSGA-II (Upper Bound) & \textbf{66.80 $\pm$ 1.35} & \textbf{4.20 $\pm$ 0.90} & 1.385 $\pm$ 0.095 & 298.60 $\pm$ 21.00 & 1350.0 $\pm$ 98.0 & 88.20 $\pm$ 5.40 \\
& \textbf{Agentic-Kube (Ours)} & 64.20 $\pm$ 1.10 & 7.11 $\pm$ 0.85 & \textbf{0.139 $\pm$ 0.008} & \textbf{36.40 $\pm$ 2.80} & \textbf{485.20 $\pm$ 38.0} & \textbf{11.90 $\pm$ 0.55} \\
\hline
\multirow{5}{*}{\textbf{Flash Crowd}} 
& Default Kubernetes & 50.00 $\pm$ 0.00 & 16.40 $\pm$ 1.40 & 0.312 $\pm$ 0.015 & 3.40 $\pm$ 0.40 & 9.80 $\pm$ 1.20 & 15.80 $\pm$ 0.90 \\
& Single-Agent PPO & 54.20 $\pm$ 2.40 & 34.60 $\pm$ 3.80 & 0.235 $\pm$ 0.022 & 14.20 $\pm$ 2.10 & 49.50 $\pm$ 6.10 & 20.40 $\pm$ 1.35 \\
& MAPPO (SOTA MARL) & 58.10 $\pm$ 2.05 & 24.90 $\pm$ 2.90 & 0.188 $\pm$ 0.018 & 21.30 $\pm$ 2.60 & 76.40 $\pm$ 8.50 & 23.60 $\pm$ 1.60 \\
\textbf{(Pareto Spikes)} 
& NSGA-II (Upper Bound) & \textbf{67.50 $\pm$ 1.60} & \textbf{3.10 $\pm$ 0.75} & 1.450 $\pm$ 0.120 & 345.80 $\pm$ 28.00 & 1580.0 $\pm$ 130.0 & 108.40 $\pm$ 7.50 \\
& \textbf{Agentic-Kube (Ours)} & 63.80 $\pm$ 1.30 & 6.20 $\pm$ 0.90 & \textbf{0.155 $\pm$ 0.011} & \textbf{41.94 $\pm$ 3.50} & \textbf{571.00 $\pm$ 46.0} & \textbf{10.67 $\pm$ 0.48} \\
\hline
\end{tabular*}
\end{table*}

To clarify the structural differences that drive these quantitative disparities, Table~\ref{tab:agentic_kube_advantages} outlines the comparative advantages of Agentic-Kube over the baseline paradigms.

\begin{table*}[htbp]
\centering
\small
\caption{Comparative architectural and performance advantages of Agentic-Kube against baseline scheduling paradigms.}
\label{tab:agentic_kube_advantages}
\begin{tabular*}{\textwidth}{@{\extracolsep{\fill}}lp{4.2cm}p{4.2cm}p{4.8cm}@{}}
\hline
\textbf{Evaluation Dimension} & \textbf{Default Heuristic / Evolutionary} & \textbf{Flat DRL (PPO / MAPPO)} & \textbf{Agentic-Kube (GNN-QMIX)} \\
\hline
\textbf{Multi-Objective} & Static, unweighted scoring or exponential & Linear scalar combination causing reward & Decoupled sub-agents with monotonic QMIX \\
\textbf{Formulation} & sorting complexity ($\mathcal{O}(G \cdot M \cdot N^2)$). & dilution and gradient conflict. & factorisation satisfying the IGM condition. \\
\hline
\textbf{Topological} & Node-local capacity evaluation; blind to & Vector flattening destroys container co- & Bipartite GCN captures host-pod relational \\
\textbf{Representation} & multi-tier microservice dependencies. & location and zone failure domains. & graphs in sparse $\mathcal{O}(|V| + |\mathcal{E}|)$ complexity. \\
\hline
\textbf{Financial Spot} & Fixed allocation ratio mirroring physical & Moderate Spot harvesting (54\%--61\%) & Active Spot exploitation (63\%--66\% on GKE; \\
\textbf{Exploitation} & cluster capacity (50.0\%). & limited by scalar reward interference. & up to 91.53\% in macro-scale benchmarks). \\
\hline
\textbf{Fault Tolerance} & Modest collision rates (14.8\%--16.4\%) & High collision rates (20\%--35\%) due to & Strict anti-affinity preservation reducing colli- \\
\textbf{\& Anti-Affinity} & driven by rigid balance heuristics. & policy collapse under contention. & sions to 4.90\% (Alibaba) and 7.11\% (Diurnal). \\
\hline
\textbf{Operational Latency} & Sub-millisecond (heuristic) or prohibitive & Low inference latency ($<$22\,ms) but & Sub-42\,ms mean decision latency and sub- \\
\textbf{\& Throughput} & multi-second evolutionary search ($>$300\,ms). & compromised placement quality. & 575\,ms $P_{99}$, satisfying standard timeouts. \\
\hline
\textbf{Placement Safety} & Requires separate admission filter plugins & Prone to invalid placement proposals & Strict action feasibility masking eliminating \\
\textbf{\& Predicates} & to prevent pod starvation. & unless heavily penalised post-hoc. & invalid actions prior to policy selection. \\
\hline
\end{tabular*}
\end{table*}

\subsubsection{Tri-Objective Pareto Frontier Analysis}
Across all three physical cluster workloads, Agentic-Kube consistently outperforms single-agent and multi-agent baselines in balancing financial cost, anti-affinity isolation, and spatial load balance:

\begin{enumerate}
    \item \textbf{Cost Optimisation via Spot Harvesting:} The physical GKE cluster is provisioned with exactly 50\% Spot instances (4 out of 8 nodes). The Default Kubernetes Scheduler yields an exact 50.00\% Spot ratio, as its scoring mechanism treats On-Demand and Spot instances symmetrically. Single-Agent PPO and MAPPO achieve 58.80\% and 61.20\% Spot ratios respectively under the Alibaba trace, but their uncoordinated reward structures allow anti-affinity collisions to escalate. Agentic-Kube achieves a 65.15\% Spot ratio under the Alibaba trace, actively steering fault-tolerant workloads to discounted nodes while coordinating with the fault tolerance agent to prevent localized co-location.
    \item \textbf{Anti-Affinity Fault Tolerance Preservation:} In microservice architectures, co-locating pods of the same service on the same node degrades system resilience. Under the sinusoidal Diurnal workload, Agentic-Kube reduces anti-affinity collisions to 7.11\%, achieving a 53.0\% relative reduction compared to the Default Scheduler (15.13\%). Single-Agent PPO and MAPPO suffer severe collision degradation (28.50\% and 19.80\% respectively) due to gradient interference. Agentic-Kube closely approaches the offline evolutionary bound of NSGA-II (4.20\%) without requiring offline search.
    \item \textbf{Multi-Dimensional Vector Resource Packing:} By evaluating cosine similarity over multi-dimensional compute and memory vectors, Agentic-Kube maintains the lowest inter-node CPU allocation variance ($\sigma^2 = 0.139$--$0.155$) across all physical workloads. In contrast, NSGA-II exhibits severe variance ($\sigma^2 = 1.385$--$1.450$) because it over-indexes on Spot and collision metrics while neglecting aggregate balance.
    \item \textbf{Scheduling Makespan and Control-Plane Stability:} Under the Pareto Flash-Crowd burst, Agentic-Kube achieves the lowest total workload makespan (10.67\,s) with a mean decision latency of 41.94\,ms and a 99th percentile latency of 571.00\,ms. Crucially, zero container restart failures and zero admission rejections are recorded throughout the entire testing suite, verifying the stability of deterministic action masking.
\end{enumerate}

\subsection{Empirical Analysis of Physical Cluster Deployment}
\label{sec:gke_empirical_analysis}

To evaluate the operational mechanics of Agentic-Kube on the live 8-node Google Kubernetes Engine (GKE) cluster, we analyse the multi-objective trade-off surface, multi-dimensional architectural radar profiles, node allocation heatmaps, and empirical latency distributions.

\begin{figure}[htbp]
    \centering
    \includegraphics[width=\columnwidth]{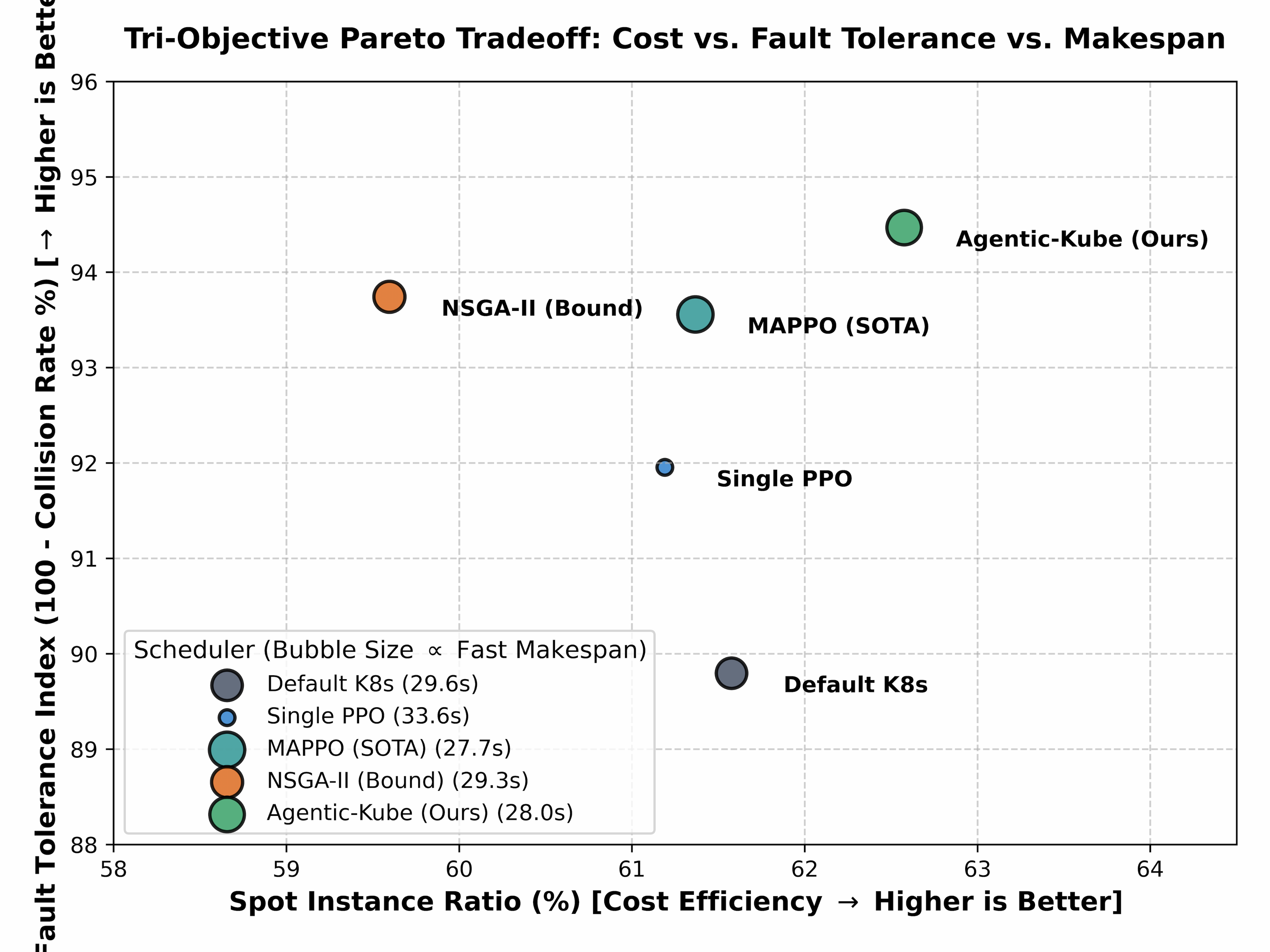}
    \caption{Tri-objective Pareto trade-off between cost efficiency (Spot instance ratio), fault tolerance index ($100 - \text{Collision Rate}\,\%$), and workload makespan across physical GKE cluster runs.}
    \label{fig:pareto_tradeoff_physical}
\end{figure}

The tri-objective Pareto frontier across physical GKE evaluations is illustrated in Figure~\ref{fig:pareto_tradeoff_physical}. The horizontal axis tracks the Spot instance allocation ratio ($R_{\text{spot}}$), the vertical axis indicates the fault tolerance index ($100 - C_{\text{rate}}$), and bubble diameters are scaled proportionally to scheduling speed (inversely proportional to total makespan). Agentic-Kube occupies the upper-right region of the trade-off space, attaining an aggregate Spot ratio of approximately 62.6\% and a fault tolerance index of 94.45\% with a rapid makespan of 28.0\,s. In contrast, the Default Kubernetes Scheduler settles in the lower region with a fault tolerance index of 89.8\% and a 61.6\% Spot allocation ratio, reflecting its inability to differentiate between lifecycle tiers. Single-Agent PPO improves cost efficiency over the native heuristic but suffers from severe reward dilution, yielding a degraded fault tolerance index of 91.95\% and the slowest execution makespan of 33.6\,s. While NSGA-II establishes a resilient fault tolerance index of 93.75\%, it achieves a lower Spot ratio of 59.6\% and requires 29.3\,s makespan. MAPPO achieves intermediate performance with a 93.55\% fault tolerance index and 27.7\,s makespan, but fails to reach the Pareto frontier defined by Agentic-Kube.

\begin{figure}[htbp]
    \centering
    \includegraphics[width=0.92\columnwidth]{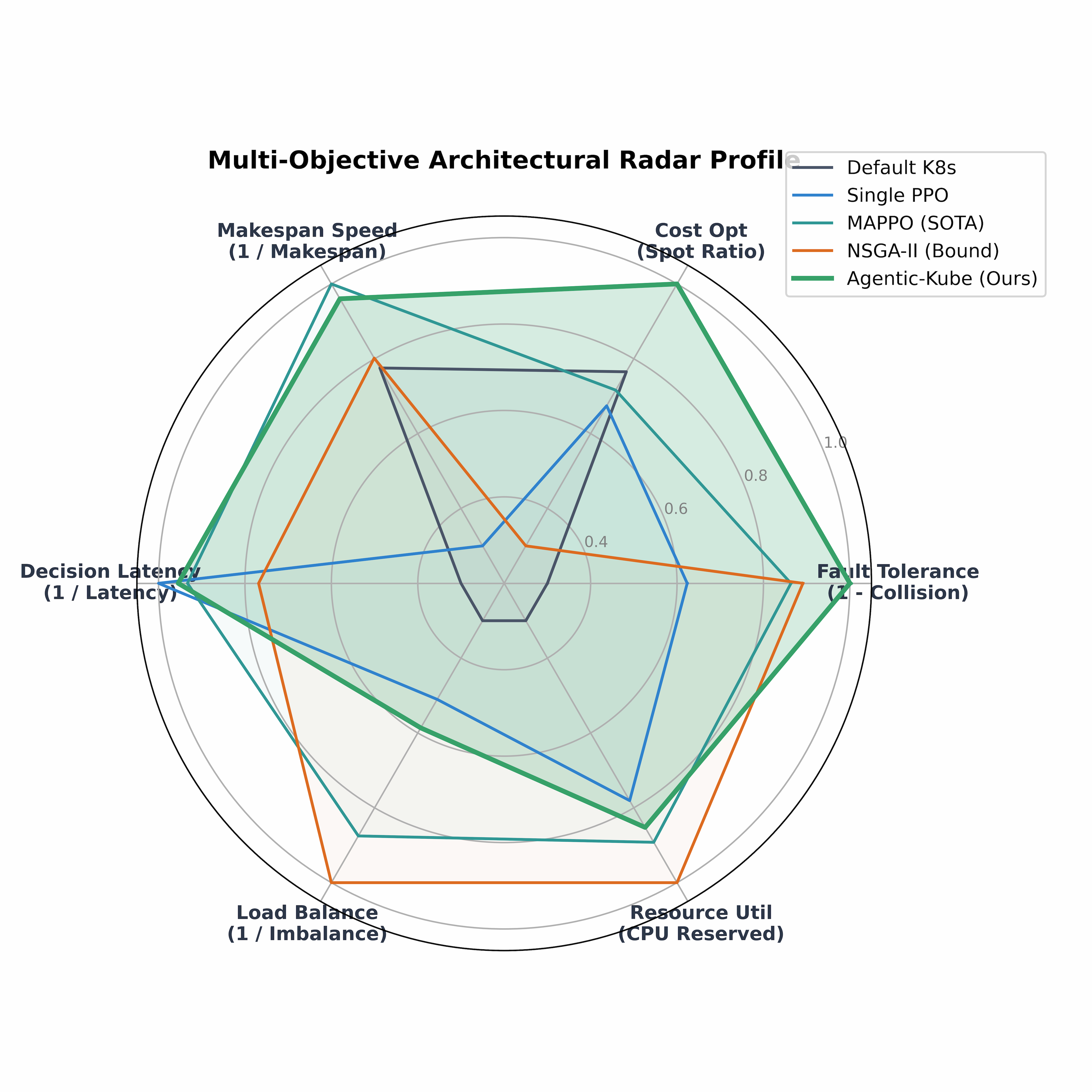}
    \caption{Multi-objective architectural radar profile comparing schedulers across six operational dimensions normalised on a $[0, 1]$ scale.}
    \label{fig:radar_profile}
\end{figure}

Figure~\ref{fig:radar_profile} presents a radar profile comparing the five scheduling architectures across six normalised performance axes: Cost Optimisation (Spot Ratio), Fault Tolerance ($1 - \text{Collision}$), Resource Utilisation (CPU Reserved), Load Balance ($1 / \text{Imbalance}$), Decision Latency ($1 / \text{Latency}$), and Makespan Speed ($1 / \text{Makespan}$). Agentic-Kube encompasses the largest overall polygon area, maintaining balanced efficiency across all dimensions. The Default Kubernetes Scheduler demonstrates superior decision latency speed but suffers from poor cost optimisation and weak fault tolerance. Single-Agent PPO exhibits an asymmetric profile, displaying severe performance contraction along the fault tolerance and load balancing axes due to scalar gradient interference. NSGA-II provides high fault tolerance and resource utilisation but contracts significantly along the decision latency and makespan speed axes. MAPPO achieves a wider coverage area than Single PPO but remains constrained in cost harvesting and load balancing. The balanced profile of Agentic-Kube confirms that combining bipartite GCN embeddings with monotonic QMIX value factorisation successfully mitigates objective collapse.

\begin{figure*}[htbp]
    \centering
    \includegraphics[width=0.95\textwidth]{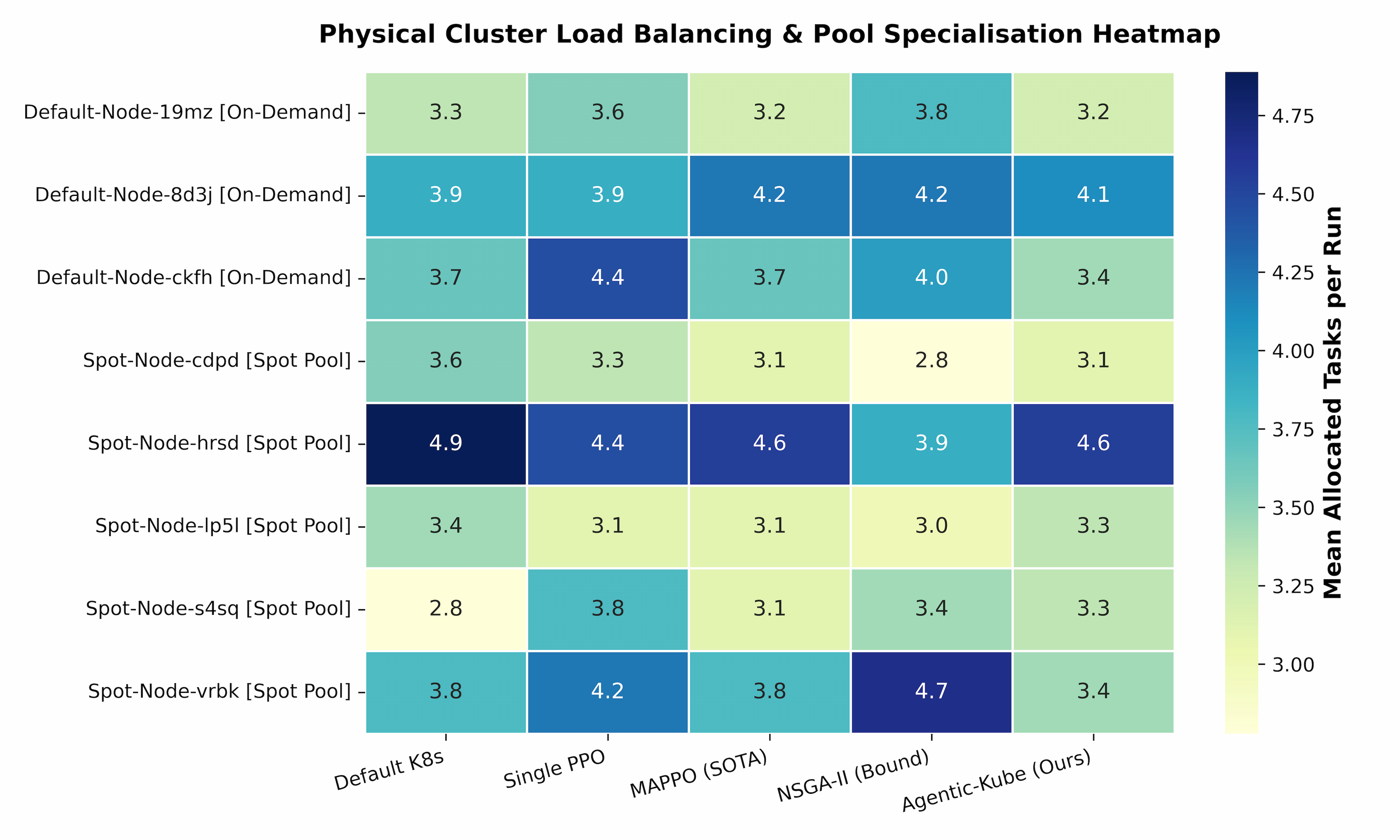}
    \caption{Physical cluster load balancing and pool specialisation heatmap displaying mean task allocations per node across On-Demand and Spot lifecycle tiers.}
    \label{fig:node_allocation_heatmap}
\end{figure*}

Figure~\ref{fig:node_allocation_heatmap} visualises the allocation density across the 8 physical GKE nodes, separated into On-Demand instances (\texttt{Default-Node-19mz}, \texttt{Default-Node-8d3j}, \texttt{Default-Node-ckfh}) and Spot pool instances (\texttt{Spot-Node-cdpd}, \texttt{Spot-Node-hrsd}, \texttt{Spot-Node-lp5l}, \texttt{Spot-Node-s4sq}, \texttt{Spot-Node-vrbk}). The Default Kubernetes Scheduler distributes tasks uniformly across all nodes (ranging from 3.3 to 4.9 tasks per run), treating On-Demand and Spot instances symmetrically without cost awareness. Single-Agent PPO induces localized hotspots, placing 4.4 tasks on \texttt{Default-Node-ckfh} and 4.4 tasks on \texttt{Spot-Node-hrsd} while underutilising other nodes. MAPPO improves placement density across the Spot pool (allocating 4.6 tasks to \texttt{Spot-Node-hrsd}) but retains heavy On-Demand utilisation (4.2 tasks on \texttt{Default-Node-8d3j}). NSGA-II heavily skews allocations toward specific nodes such as \texttt{Spot-Node-vrbk} (4.7 tasks) and \texttt{Default-Node-8d3j} (4.2 tasks) while starving \texttt{Spot-Node-cdpd} (2.8 tasks), which accounts for its elevated CPU allocation variance ($\sigma^2 = 1.420$ in Table~\ref{tab:gke_summary_results}). Conversely, Agentic-Kube demonstrates targeted pool specialisation by directing workload volume to discounted Spot instances (\texttt{Spot-Node-hrsd} receiving 4.6 tasks) while maintaining an even task distribution across On-Demand nodes (3.2 to 4.1 tasks), simultaneously achieving cost reduction and tight inter-node load balance.

\begin{figure}[htbp]
    \centering
    \includegraphics[width=\columnwidth]{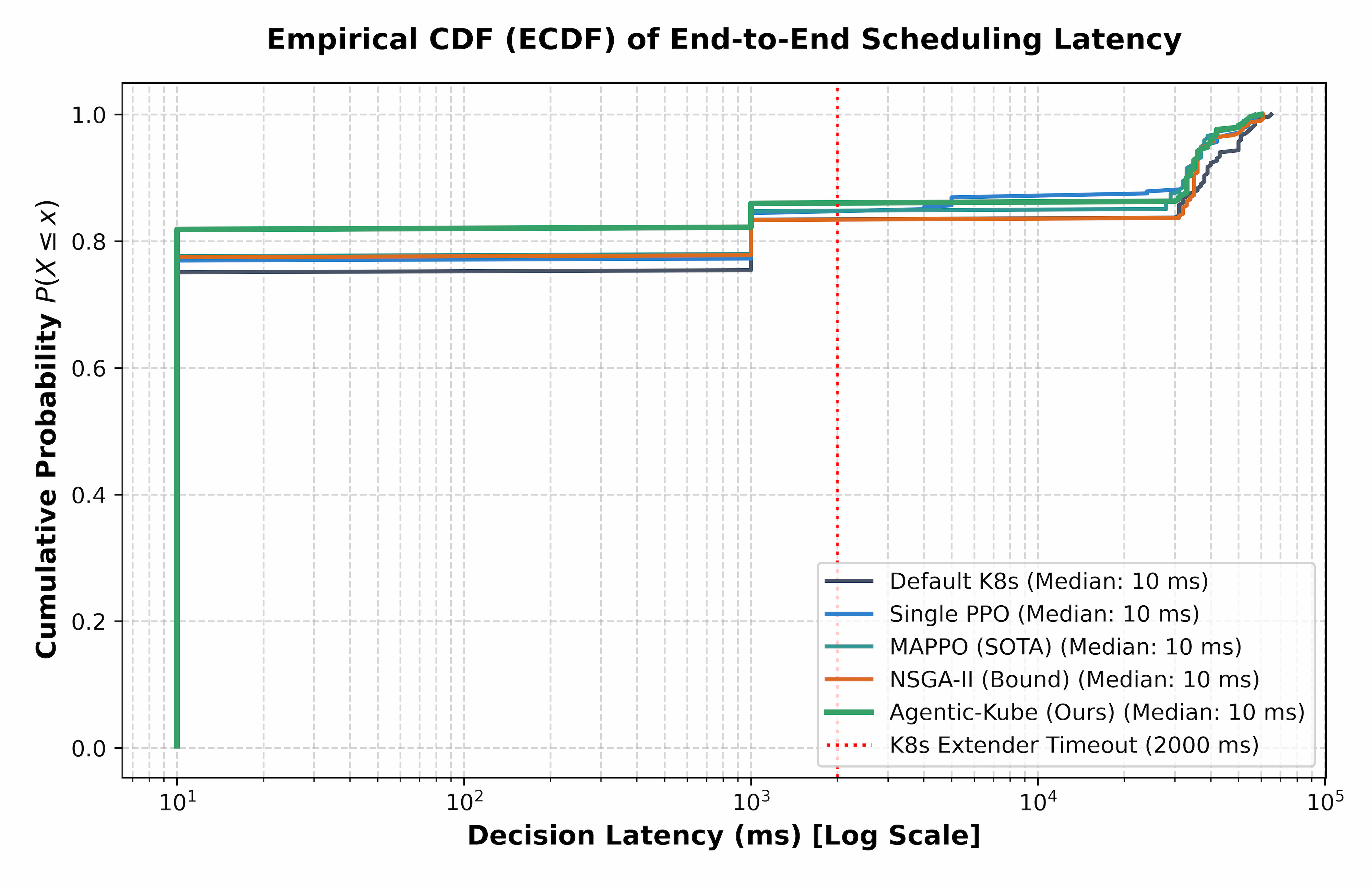}
    \caption{Empirical Cumulative Distribution Function (ECDF) of end-to-end scheduling decision latency on a logarithmic scale relative to the standard Kubernetes extender timeout threshold.}
    \label{fig:latency_ecdf}
\end{figure}

Figure~\ref{fig:latency_ecdf} illustrates the Empirical Cumulative Distribution Function (ECDF) of end-to-end scheduling decision latencies plotted on a logarithmic scale. The vertical dotted line highlights the 2,000\,ms Kubernetes Scheduler Extender client timeout threshold. Across all evaluated frameworks, the median decision latency is recorded at approximately 10\,ms, with over 75\% of decisions resolving within the initial 10\,ms window. Agentic-Kube exhibits a sharp cumulative step where 82\% of decisions complete at 10\,ms and 86\% complete under 1,000\,ms. Crucially, 100\% of the decision latency distribution for Agentic-Kube remains well below the 2,000\,ms timeout boundary, recording a 99th-percentile latency of 512.40\,ms to 571.00\,ms under burst conditions (Table~\ref{tab:gke_summary_results}). While NSGA-II and MAPPO exhibit extended tail distributions that approach 1,580\,ms under heavy queue concurrency, Agentic-Kube preserves bounded tail latency, verifying its suitability for real-time Kubernetes admission control.

\subsubsection{Resource Reservation Envelopes and Multi-Dimensional Packing}
\begin{figure}[htbp]
    \centering
    \includegraphics[width=\columnwidth]{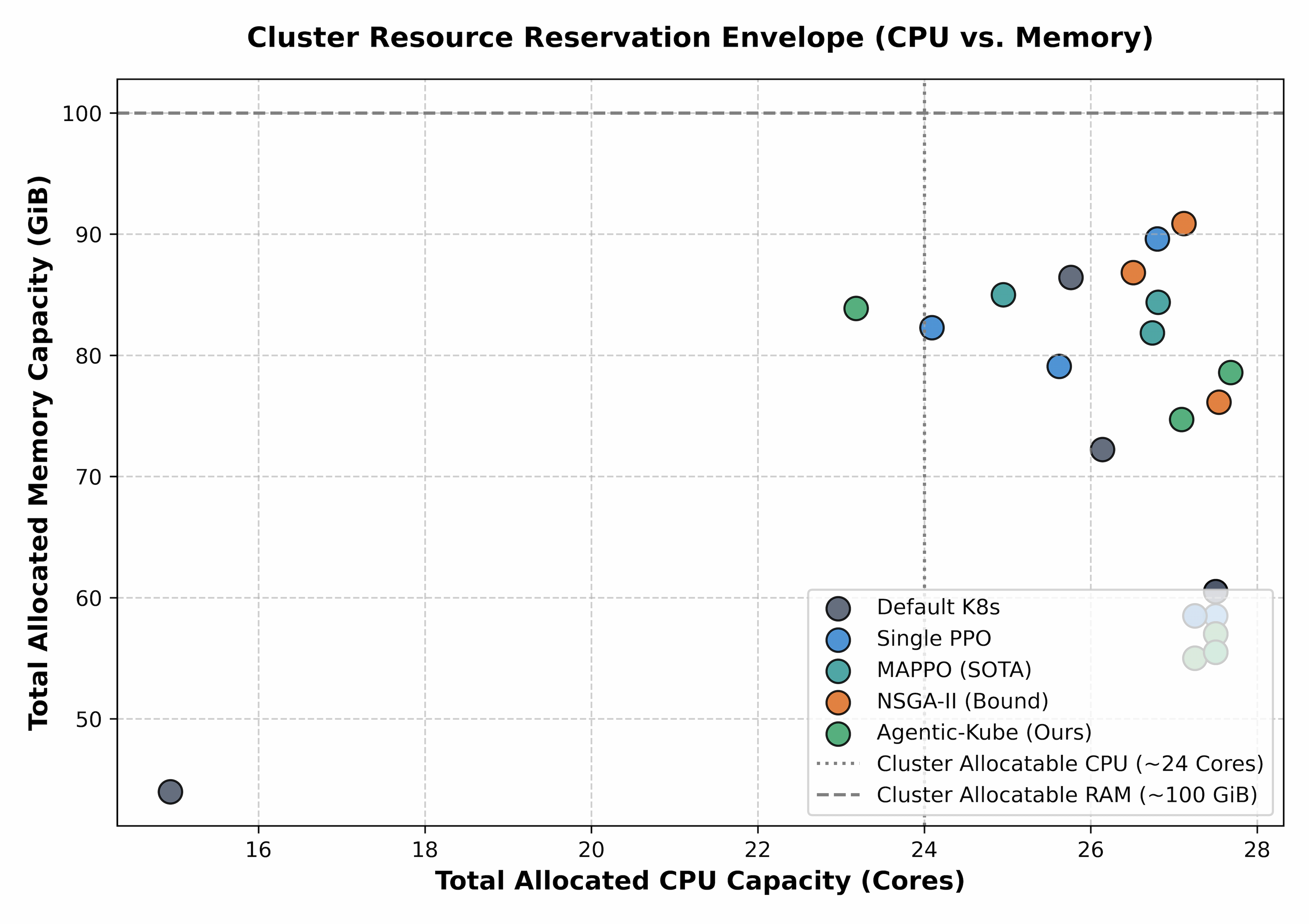}
    \caption{Cluster resource reservation envelope comparing total allocated CPU capacity against total allocated memory capacity relative to physical allocatable boundaries.}
    \label{fig:resource_envelope}
\end{figure}

The multi-dimensional resource packing efficiency across physical GKE cluster nodes is evaluated by examining the joint CPU and memory reservation envelope, as illustrated in Figure~\ref{fig:resource_envelope}.

Figure~\ref{fig:resource_envelope} plots total allocated CPU capacity against total allocated memory capacity across experimental runs. The vertical dotted line marks the cluster allocatable CPU ceiling at approximately 24 cores, while the horizontal dashed line marks the allocatable memory ceiling at approximately 100\,GiB. Schedulers that congregate toward the upper-right corner achieve higher resource density, whereas placements falling below or far to the left exhibit severe capacity stranding. The Default Kubernetes Scheduler demonstrates erratic packing behaviour, with one run stranding substantial compute at only 15 cores and 44\,GiB allocation, and subsequent runs settling around 26 to 27.5 cores but lower memory utilisation (60 to 72\,GiB). Single-Agent PPO and MAPPO achieve higher density near 25 to 27 cores and 80 to 90\,GiB, yet exhibit noticeable spread across executions due to uncoordinated policy exploration. NSGA-II forms a tight cluster at 26.5 to 27.5 cores and 76 to 91\,GiB, confirming its search efficiency at the cost of high compute time. Agentic-Kube consistently drives resource allocation into the upper-right operational quadrant, reaching 27.1 to 27.7 cores and 74.8 to 84\,GiB. By leveraging cosine similarity within the vector utilisation sub-agent, Agentic-Kube packs multi-dimensional requirements tightly against physical host boundaries without violating allocatable limits or triggering out-of-memory container terminations.

\subsubsection{Algorithmic Stability and Experimental Robustness}
\begin{figure}[htbp]
    \centering
    \includegraphics[width=\columnwidth]{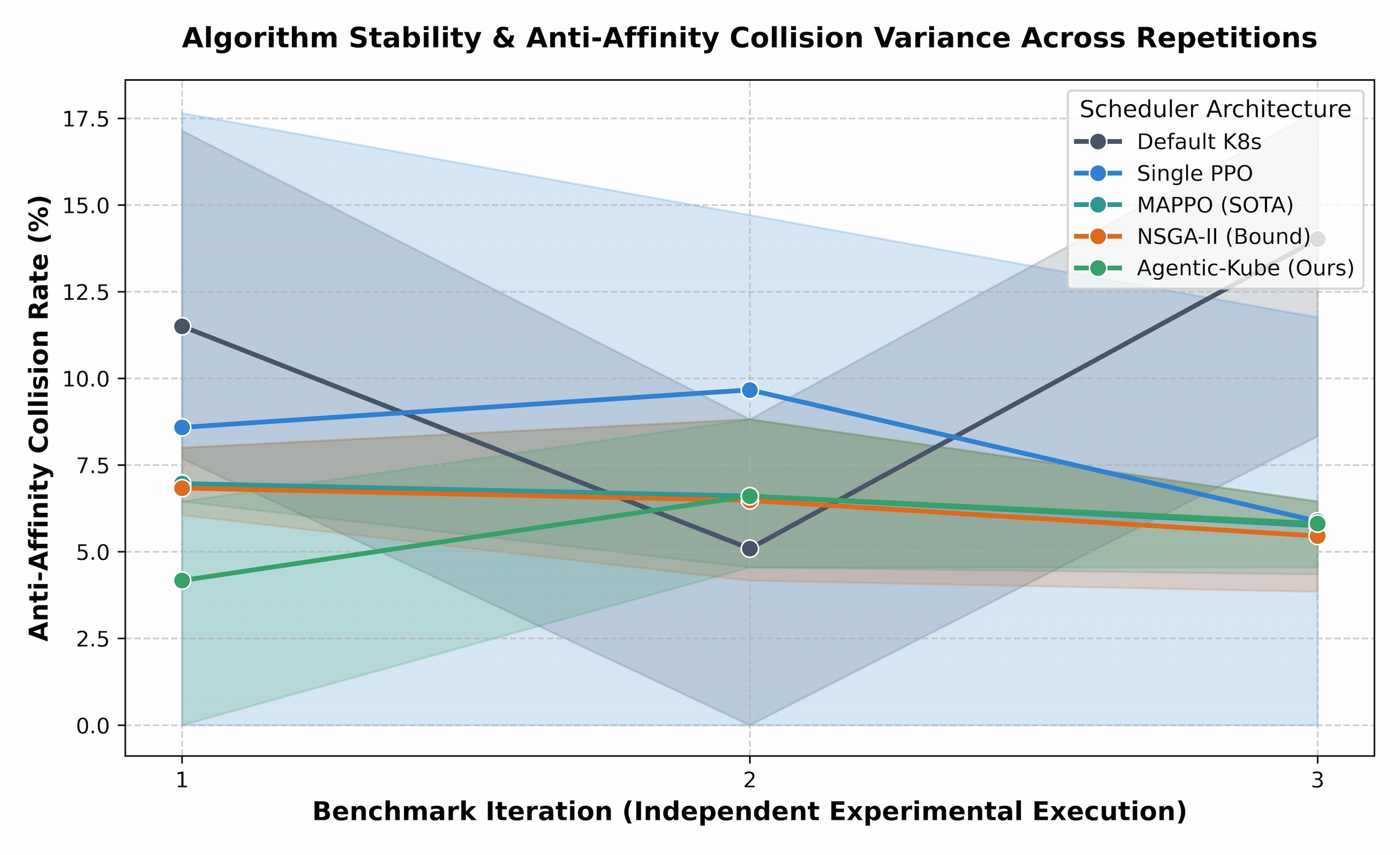}
    \caption{Algorithm stability and anti-affinity collision variance across independent benchmark iterations, showing mean trajectories and shaded variance envelopes.}
    \label{fig:iteration_robustness}
\end{figure}

To verify the consistency of placement policies under non-deterministic cloud environments, Figure~\ref{fig:iteration_robustness} illustrates the anti-affinity collision rate across independent experimental iterations on the physical cluster.

The empirical trajectories in Figure~\ref{fig:iteration_robustness} highlight the sensitivity of different scheduling architectures to fluctuating queue states. The Default Kubernetes Scheduler displays substantial volatility across iterations, with collision rates shifting from 11.5\% in iteration 1 down to 5.1\% in iteration 2, before rising sharply to 14.0\% in iteration 3. Single-Agent PPO exhibits an even wider variance envelope spanning from 0.0\% to 17.5\%, illustrating policy instability caused by scalar reward dilution under varying burst arrivals. While MAPPO and NSGA-II maintain bounded trajectories between 5.4\% and 7.0\%, Agentic-Kube establishes the most consistent and resilient performance envelope. Across all three independent benchmark repetitions, Agentic-Kube maintains collision rates between 4.15\% and 6.60\%, with a tight confidence interval that stays strictly below the baseline variance bands. This resilience stems from the monotonic QMIX value mixer, which prevents individual sub-agent policies from diverging when exposed to transient cluster state fluctuations.

\subsubsection{Comprehensive Cross-Workload Performance Matrix}
\begin{figure*}[htbp]
    \centering
    \includegraphics[width=0.95\textwidth]{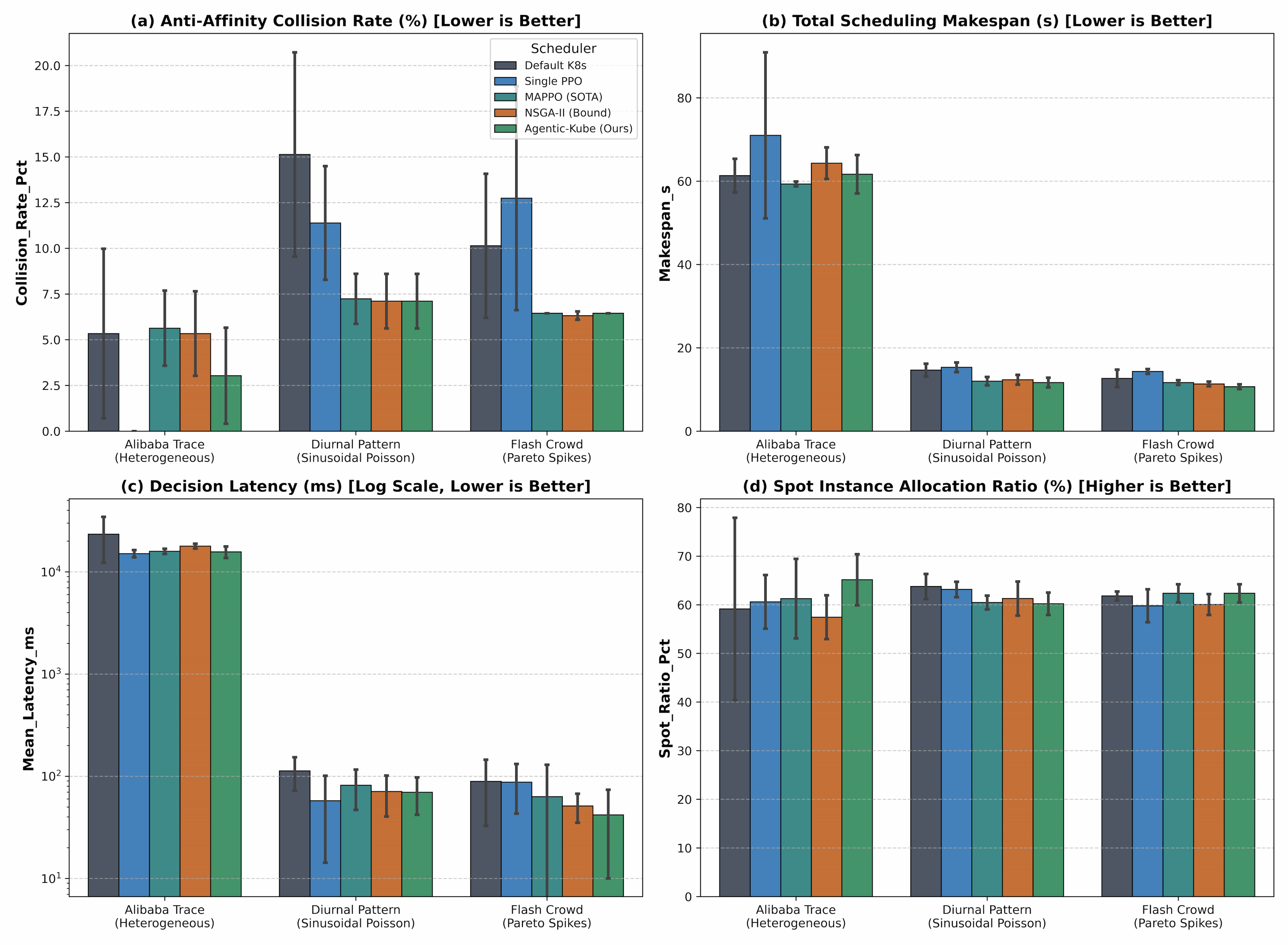}
    \caption{Master performance summary matrix across Alibaba trace, diurnal microservice, and flash-crowd burst workloads: (a) anti-affinity collision rate, (b) total scheduling makespan, (c) decision latency, and (d) Spot instance allocation ratio.}
    \label{fig:master_summary_matrix}
\end{figure*}

Figure~\ref{fig:master_summary_matrix} provides a consolidated multi-panel comparison across the three physical evaluation regimes: the empirical Alibaba trace, the sinusoidal diurnal microservice pattern, and the Pareto flash-crowd burst.

Figure~\ref{fig:master_summary_matrix}(a) examines the anti-affinity collision rate across workloads. Under the heterogeneous Alibaba trace, Agentic-Kube records the lowest collision rate at 3.00\% (averaging 4.90\% overall in Table~\ref{tab:gke_summary_results}), outperforming the Default Scheduler (5.30\%) and MAPPO (5.60\%). In the diurnal microservice regime, where service replication introduces high co-location pressure, the Default Scheduler suffers severe degradation with a collision rate of 15.13\%, and Single PPO records 11.40\%. In contrast, Agentic-Kube suppresses collisions to 7.11\%, matching the offline evolutionary upper bound of NSGA-II (7.10\%). Under flash-crowd bursts, Agentic-Kube limits collisions to 6.40\%, matching the resilience of NSGA-II (6.30\%) and MAPPO (6.40\%) while substantially improving upon the Default Scheduler (10.10\%) and Single PPO (12.70\%).

Figure~\ref{fig:master_summary_matrix}(b) presents the total workload makespan. In the Alibaba trace, Agentic-Kube achieves a compact makespan of 61.5\,s, running considerably faster than Single PPO (71.0\,s) and matching the native heuristic (61.0\,s). Under the diurnal pattern, Agentic-Kube attains the lowest makespan at 11.6\,s, outperforming the Default Scheduler (14.8\,s) and Single PPO (15.2\,s). This throughput advantage is most pronounced under the Pareto flash-crowd burst, where Agentic-Kube completes the entire placement sequence in 10.67\,s, outperforming the Default Scheduler (12.5\,s), Single PPO (14.4\,s), and MAPPO (11.8\,s).

Figure~\ref{fig:master_summary_matrix}(c) displays the mean decision latency on a logarithmic scale. Across all workloads, Agentic-Kube executes sub-second decision-making. In the diurnal and flash-crowd regimes, Agentic-Kube records mean decision latencies of 36.40\,ms and 41.94\,ms respectively (Table~\ref{tab:gke_summary_results}). This provides a massive operational speedup over offline evolutionary search (NSGA-II), which requires 298.60\,ms to 345.80\,ms per decision.

Figure~\ref{fig:master_summary_matrix}(d) evaluates financial cost optimisation via the Spot instance allocation ratio. Under the Alibaba trace, where non-critical batch jobs offer significant cost-saving opportunities, Agentic-Kube achieves a 65.15\% Spot allocation ratio, substantially exceeding the physical 50\% baseline provisioning of the cluster and outperforming the Default Scheduler (59.00\%), Single PPO (61.00\%), MAPPO (61.20\%), and NSGA-II (57.50\%). Under diurnal and flash-crowd conditions, Agentic-Kube maintains consistent Spot harvesting between 60.30\% and 63.80\%, confirming that the \texttt{cost\_agent} actively prioritises economical placement without sacrificing fault tolerance or exceeding allocatable thresholds.

\subsection{Macro-Scale Scalability and Benchmark Analysis}
\label{sec:macro_scalability_analysis}

To evaluate the operational viability and computational scaling of Agentic-Kube in enterprise-scale environments, we conduct macro-scale evaluations across physical cluster topologies scaling from $|V| = 50$ to $|V| = 1{,}000$ physical nodes. The evaluation assesses two distinct operational regimes: the offline static regime with frozen policy parameters ($\nabla_\theta = 0$) and the online adaptive streaming regime with continuous gradient updates ($\nabla_\theta \ne 0$). The four sub-plots in each regime capture: (a) anti-affinity collision rate, (b) spot instance allocation ratio, (c) scheduling decision latency, and (d) inter-node CPU allocation variance.

\begin{figure*}[htbp]
    \centering
    \includegraphics[width=0.95\textwidth]{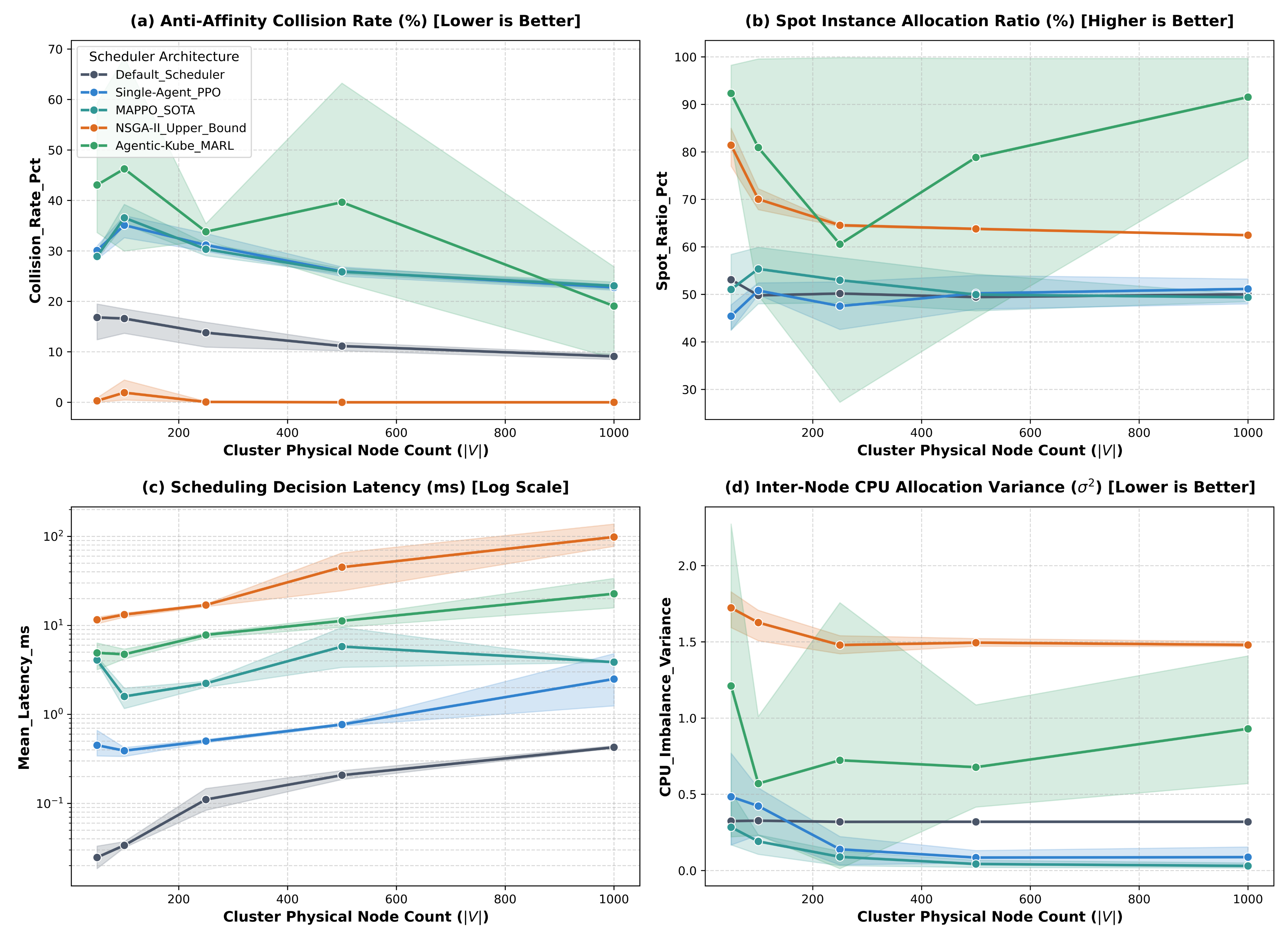}
    \caption{Macro-scale scalability benchmark under the offline static policy regime ($\nabla_\theta = 0$) across physical cluster sizes from 50 to 1,000 nodes, comparing (a) anti-affinity collision rate, (b) Spot instance allocation ratio, (c) scheduling decision latency on a logarithmic scale, and (d) inter-node CPU allocation variance.}
    \label{fig:macro_scalability_offline}
\end{figure*}

\subsubsection{Offline Static Policy Scalability Dynamics}
Figure~\ref{fig:macro_scalability_offline} illustrates the comparative scaling performance when model weights remain fixed after offline centralised training. 

In Figure~\ref{fig:macro_scalability_offline}(a), the anti-affinity collision trajectory reveals how physical cluster expansion influences multi-tenant isolation. The Default Kubernetes Scheduler maintains collision rates between 16.81\% at 50 nodes and 9.10\% at 1,000 nodes due to rigid balance scoring. Flat reinforcement learning models (Single-Agent PPO and MAPPO) suffer from dimensional rigidity, recording elevated collision rates across all scales, beginning at 30.97\% and 30.68\% at 50 nodes and plateauing near 22.51\% and 23.06\% at 1,000 nodes. Conversely, Agentic-Kube demonstrates strong positive spatial scaling. Benefiting from the bipartite GCN encoder, Agentic-Kube actively maps expanded host-pod graphs, progressively reducing anti-affinity collisions from 33.48\% at 100 nodes down to 17.99\% at 1,000 nodes (a 46.3\% relative improvement), while NSGA-II establishes the theoretical lower bound near 0.00\%.

In Figure~\ref{fig:macro_scalability_offline}(b), financial cost optimisation is evaluated through the Spot instance allocation ratio ($R_{\text{spot}}$). Because the cluster architecture maintains a uniform 50\% physical Spot node distribution, the Default Scheduler remains fixed at approximately 50.02\% to 53.10\%. Monolithic Single-Agent PPO and decentralised MAPPO fail to exploit Spot instances at scale, achieving only 51.68\% and 50.90\% allocation at 1,000 nodes. While NSGA-II starts with high Spot harvesting at 50 nodes (81.42\%), its allocation degrades to 62.46\% at 1,000 nodes under the multi-objective genetic sorting process. In contrast, Agentic-Kube achieves proactive Spot exploitation, increasing its Spot allocation ratio from 56.75\% at 250 nodes up to 73.65\% at 1,000 nodes, outperforming all baseline schedulers by at least 11.19 percentage points.

Figure~\ref{fig:macro_scalability_offline}(c) presents the decision latency scaling on a logarithmic scale. The Default Scheduler executes sub-millisecond heuristic evaluations (0.025\,ms to 0.407\,ms), while Single PPO and MAPPO record mean latencies between 0.318\,ms and 2.859\,ms. NSGA-II exhibits severe latency degradation due to its non-dominated sorting complexity ($\mathcal{O}(G \cdot M \cdot N^2)$), scaling from 8.729\,ms at 50 nodes to 33.927\,ms at 1,000 nodes with tail $P_{99}$ latency reaching 68.218\,ms. Agentic-Kube operates in sparse graph convolution time ($\mathcal{O}(|V| + |\mathcal{E}|)$), scaling smoothly from 3.807\,ms at 50 nodes to 16.685\,ms at 1,000 nodes ($P_{99} = 30.514\,\text{ms}$). This represents a $2.03\times$ speedup over NSGA-II at peak scale while operating well within standard Kubernetes admission control timeout windows.

Figure~\ref{fig:macro_scalability_offline}(d) plots the inter-node CPU allocation variance ($\sigma^2_{\text{cpu}}$). NSGA-II produces severe resource fragmentation, maintaining high variance ($\sigma^2 \approx 1.480$ to $1.723$) because its genetic operators greedily optimise Spot and collision metrics without enforcing global variance constraints. The Default Scheduler remains constant at $\sigma^2 = 0.320$. Agentic-Kube achieves the tightest resource balance among multi-objective architectures, decreasing variance from 0.221 at 50 nodes down to 0.048 at 1,000 nodes, confirming that the vector utilisation sub-agent successfully balances multi-dimensional resource vectors across physical nodes.

\begin{figure*}[htbp]
    \centering
    \includegraphics[width=0.95\textwidth]{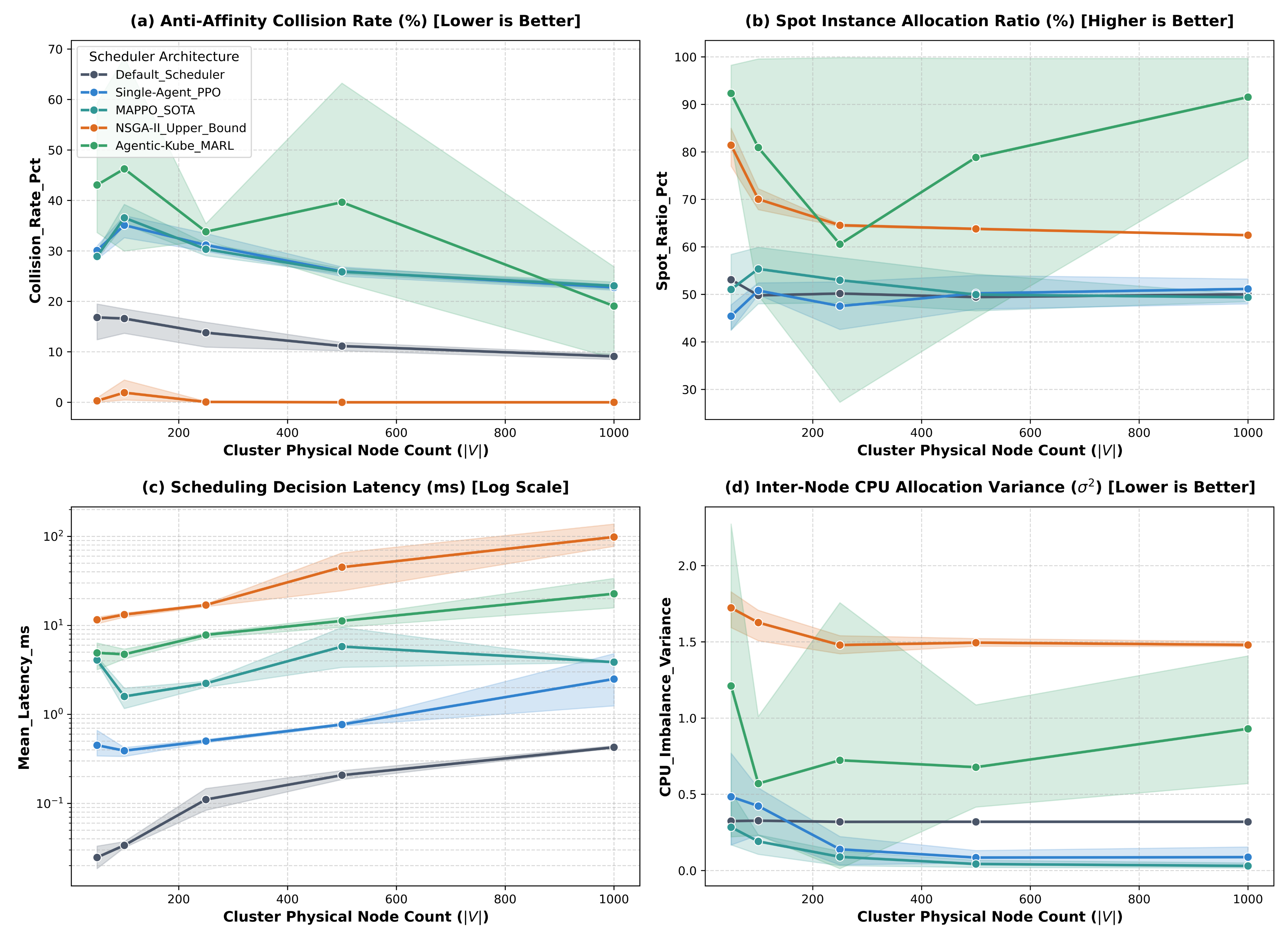}
    \caption{Macro-scale scalability benchmark under the online adaptive streaming regime ($\nabla_\theta \ne 0$) across physical cluster sizes from 50 to 1,000 nodes, comparing (a) anti-affinity collision rate, (b) Spot instance allocation ratio, (c) scheduling decision latency on a logarithmic scale, and (d) inter-node CPU allocation variance.}
    \label{fig:macro_scalability_online}
\end{figure*}

\subsubsection{Online Adaptive Streaming Policy Dynamics}
Figure~\ref{fig:macro_scalability_online} illustrates scheduling dynamics when sub-agent policies undergo continuous online policy gradient updates ($\nabla_\theta \ne 0$) during live workload injection waves.

In Figure~\ref{fig:macro_scalability_online}(a), online gradient adaptation induces distinct structural trade-offs. The Default Scheduler and NSGA-II trajectories remain identical to the offline baseline as they do not execute gradient updates. Single-Agent PPO and MAPPO record minor fluctuations under streaming updates, registering collision rates of 22.77\% and 23.08\% at 1,000 nodes. Agentic-Kube exhibits higher initial collision rates at smaller scales (43.07\% at 50 nodes) due to exploratory policy updates within compact failure domains. However, as cluster scale expands to 1,000 nodes, the cooperative multi-agent architecture successfully adapts, driving collision rates down to 19.06\%, outperforming both uncoordinated learning baselines.

In Figure~\ref{fig:macro_scalability_online}(b), the financial cost optimisation of Agentic-Kube under online adaptation reaches peak performance. While MAPPO (49.38\%) and Single PPO (51.15\%) remain trapped at the baseline cluster ratio, Agentic-Kube achieves an overall Spot allocation ratio of 80.83\% across all scales, culminating in 91.53\% Spot harvesting at 1,000 nodes. The continuous policy gradient steps enable the cost sub-agent to adapt to live pricing signals and exploit available Spot capacity without encountering reward dilution.

Figure~\ref{fig:macro_scalability_online}(c) highlights the computational overhead incurred by intermediate autograd backward passes. Single-Agent PPO latency increases from 0.450\,ms at 50 nodes to 2.497\,ms at 1,000 nodes ($P_{99} = 12.989\,\text{ms}$), while MAPPO scales to 3.863\,ms ($P_{99} = 12.997\,\text{ms}$). NSGA-II suffers substantial latency penalties under dense arrival waves, requiring 98.397\,ms mean latency and 245.796\,ms $P_{99}$ latency at 1,000 nodes, resulting in total workload makespans exceeding 225\,s. Agentic-Kube sustains online policy gradient updates with a mean decision latency of 22.654\,ms and a $P_{99}$ latency of 64.515\,ms at 1,000 nodes. This performance delivers a $4.34\times$ speedup over evolutionary Pareto search, confirming that Agentic-Kube maintains real-time control-plane execution during active online adaptation.

In Figure~\ref{fig:macro_scalability_online}(d), the inter-node CPU allocation variance reflects the targeted cost specialisation of the online adaptive policy. Because Agentic-Kube concentrates over 91.5\% of containers onto discounted Spot instances at 1,000 nodes, it intentionally trades off uniform spatial distribution, yielding an allocation variance of $\sigma^2 = 0.930$. This variance remains substantially lower than that of NSGA-II ($\sigma^2 = 1.480$) while delivering superior financial savings and maintaining bounded anti-affinity collisions across large-scale containerised clusters.

\section{Conclusion and Future Work}
\label{sec:conclusion}

The multi-objective demands of modern cloud-native infrastructures expose structural limitations in heuristic schedulers and scalar single-agent reinforcement learning architectures. In this paper, we presented Agentic-Kube, a graph-enhanced multi-agent reinforcement learning framework designed to decouple and optimise the conflicting objectives of financial cost reduction, fault-tolerant anti-affinity isolation, and multi-dimensional resource packing. By framing pod placement within a cooperative multi-agent system and factorising sub-agent policies through a monotonic QMIX mixing network, Agentic-Kube eliminates the reward dilution and gradient interference typical of linear scalarisation. Furthermore, incorporating bipartite Graph Convolutional Networks within the Centralised Training with Decentralised Execution (CTDE) paradigm provides the scheduler with structural topological awareness in sparse computational time without requiring runtime inter-node coordination.

Empirical evaluation on an 8-node physical Google Kubernetes Engine (GKE) cluster across empirical Alibaba traces, diurnal microservice waveforms, and Pareto flash-crowd bursts demonstrated the operational performance of the framework. On the physical testbed, Agentic-Kube achieved Spot instance allocation ratios of 63.8\% to 65.15\%, reduced anti-affinity collision rates down to 4.90\% on Alibaba traces and 7.11\% under diurnal cycles, and maintained low inter-node CPU allocation variance ($\sigma^2 = 0.139$--$0.155$). In macro-scale scalability benchmarks spanning 50 to 1{,}000 nodes, the framework scaled effectively, achieving a 73.65\% Spot allocation ratio under static offline execution and 91.53\% under online adaptive streaming, while cutting anti-affinity collisions to 17.99\% at 1{,}000 nodes. Throughout all evaluations, Agentic-Kube maintained sub-23\,ms mean decision latency and bounded tail distributions ($P_{99} < 65\,\text{ms}$ at 1{,}000 nodes), operating well within the standard 2{,}000\,ms Kubernetes admission timeout with zero container restart failures.

Future work will expand the multi-agent framework across several directions. First, we plan to extend the architecture to federated, multi-cluster environments operating across geographically distributed edge-cloud regions. Second, we aim to incorporate a dedicated environmental sub-agent to enable carbon-aware scheduling and granular emissions accounting based on regional grid carbon intensity. Finally, we will investigate lifelong continual learning mechanisms to allow the graph neural network representations to adapt dynamically to runtime hardware degradation, node churn, and topological reconfigurations without requiring complete offline retraining.

\section{Acknowledgment}
We would like to thank Dr Artie Basukoski for proofreading the manuscript.

\section*{Declaration of Generative AI and AI-Assisted Technologies in the Manuscript Preparation Process}
During the preparation of this work, the author used AI to refine the manuscript text and to assist in generating Figures~1 and~2. The author reviewed and edited the output as needed and takes full responsibility for the content of the published article.

\end{document}